\documentclass[aps,prd,groupedaddress]{revtex4-1}
\usepackage{amsmath,amssymb,amsthm}
\usepackage{graphicx}
\usepackage{float}
\usepackage{color}

\newcommand{\be}{\begin{equation}}
\newcommand{\ee}{\end{equation}}
\newcommand{\bea}{\begin{array}}
\newcommand{\ea}{\end{array}}
\newcommand{\beqa}{\begin{eqnarray}}
\newcommand{\eeqa}{\end{eqnarray}}
\newcommand{\bean}{\begin{eqnarray*}}
\newcommand{\eean}{\end{eqnarray*}}

\def\up#1{\leavevmode \raise.16ex\hbox{#1}}

\newcommand{\gapproxeq}{\lower
 .7ex\hbox{$\;\stackrel{\textstyle >}{\sim}\;$}}
\newcommand{\lapproxeq}{\lower .7ex\hbox{$\;\stackrel
{\textstyle <}{\sim}\;$}}

\newcounter{appendice}

\begin{document}
\begin{flushright}
ICN 2014-01-10 \\
\end{flushright}

\title{A Twisted ${\mathcal C}^{\star}$ - algebra
formulation of Quantum Cosmology \\
with application to the Bianchi I
model}

\author{Marcos Rosenbaum}
 \email{mrosen@nucleares.unam.mx}
\author{J. David Vergara}%
 \email{vergara@nucleares.unam.mx}
\author{Rom\'an Ju\'arez}
 \email{roman.juarez@nucleares.unam.mx}
\affiliation{%
 Instituto de Ciencias Nucleares, Universidad Nacional Aut\'onoma de M\'exico,\\
 A. Postal 70-543 , M\'exico D.F., M\'exico
}%

\author{A.A. Minzoni}
\email{tim@mym.iimas.unam.mx} \affiliation{Instituto de
Investigaci\'on en Matem\'aticas Aplicadas
y en Sistemas,\\
 Universidad Nacional Aut\'onoma de M\'exico,\\
 A. Postal 70-543 , M\'exico D.F., M\'exico
}%

\date{\today}

\begin{abstract}
A twisted ${\mathcal C}^\star $- algebra of the extended
(noncommutative) Heisenberg-Weyl group has been constructed which
takes into account the Uncertainty Principle for coordinates in the
Planck length regime. This general construction is then used to
generate an appropriate Hilbert space and observables for the
noncommutative theory which, when applied to the Bianchi I
Cosmology, leads to a new set of equations that describe the quantum
evolution of the universe. We find that this formulation matches
theories based on a reticular Heisenberg-Weyl algebra in the
bouncing and expanding regions of a collapsing Bianchi universe.
There is, however, an additional effect introduced by the dynamics
generated by the noncommutativity. This is an oscillation in the
spectrum of the volume operator of the universe, within the bouncing
region of the commutative theories. We show that this effect is
generic and produced by the noncommutative momentum exchange between
the degrees of freedom in the cosmology. We give asymptotic and
numerical solutions which show the above mentioned effects of the
noncommutativity.
\begin{description}
\item[PACS:03.70.+k, 98.80.Qc, 04.60.Pp]
\end{description}
\end{abstract}

\maketitle

\section{Introduction}
Reductionism  is an essential concept in Physics which
has been validated by experiments involving energies ranging from
orders of $eV$'s in molecular and atomic physics to a few $TeV$ in
the strong interaction regime. This paradigm has led to such
successes of
quantum unification as the Standard Model, involving Electromagnetic, Weak and Strong Interactions. However the oldest interaction known to man: Gravity, and its most beautiful geometrical formulation: General Relativity, have to this day avoided quantization and even more so, unification with the other three fundamental forces of Nature.  Thus Quantization of Relativity at distances of the order of the Planck length and energies of the order of $10^{16} TeV$, still remains to be one of the most compelling problems in the field, mainly due to the lack of experimental data that could help shed some more light on which path should one pursue. \\

Because Quantum Cosmology can be seen as a minisuperspace of Quantum Gravity  where most of the degrees of freedom have been frozen and, although there is no {\it a priori}  reason to assume that the conclusions derived from the former can be readily translated to the later, it is expected that some approaches to Quantum Cosmology
can provide a convenient initial framework to investigate quantum processes involving distances of the order of Planck lengths where manifestations of noncommutativity
should occur.\\
The main purpose of this paper is to provide what we consider might be one such self-consistent formulation for Quantum
Cosmology that could lead to further insights and directives towards Quantum Gravity at scales where the implications
of the Uncertainty Principle of Quantum Mechanics and the Principle of Equivalence of Gravitation become commensurate.

Indeed, regardless of which will be eventually the final and
complete Theory for Quantum Gravity, it seems that the present
attempts for its formulation have as a common denominator some
concept of noncommutativity ( {\it see e.g.} \cite{dop},
\cite{fred}, \cite{rob}, \cite{camelia}, \cite{tomassi},
\cite{martinetti} ). Thus, in addition to the fact that Physics is a
discipline based on experiment and that a theory needs to be
validated or dismissed only on this basis before its ultimate
acceptance, it is sensible to expect that the concept of
noncommutativity should be a self-consistent part of it. One
formulation that appeals to many physicists in the field is String
Theory \cite{CDS}. Several research groups in Relativity on the
other hand believe that a more geometrical approach such as Loop
Quantum Gravity (LQG) constitutes an equally viable candidate (see
{\it e.g.} \cite{ash0}) and, on the other extreme of the theory
spectrum, is the Noncommutative Geometry developed by A. Connes and
others
(see {\it e.g.} \cite{connes1}, \cite{cori}, \cite{landi}, \cite{mas-mar}). \\
As pointed out in the Review by Douglas and Nekrasov \cite{doune},
some of the strong arguments in favor of noncommutativity and of
further support for Noncommutative Geometry originated from these
varied approaches has led to a flurry of activities and trends where
mathematical clarity and conceptual self-consistency "appear less
central to physical considerations". Examples of such a case are the
earlier quantum cosmology formulations based on a Bopp map
deformation of the Wheeler-De Witt equation, resulting from
inserting a Moyal $\star$-product between the classical Hamiltonian
and the elements of the Hilbert vector space of wave functions.
This, from the viewpoint of Deformation Quantization where the Moyal
$\star$-product arises as a deformation of the algebra product of
the Weyl symbols of quantum operator observables, has no conceptual
support. Moreover, as we have shown in \cite{rm} (and references
therein) a more logical noncommutative replacement for the
Schr\"odinger equation is the $\star$-value equation involving the
deformed Moyal $\star$-product of the Weyl symbol of the quantum
Hamiltonian operator and the Wigner function. It may be meaningful
to notice here also that in a previous work \cite{VKS}  of the type
mentioned above, the region close to the singularity has not been
explored and the wave functions have branch points which imply an
undetermined behavior near the singularity, which could very well be
attributed to the authors use of this unsubstantiated Moyal product
in the Wheeler-de Witt equation.

Alternatively, the ${\mathcal C}^\star$-algebra $\frak A$, on which our approach is based, is in particular a good example of the strategy of Noncommutative Geometry, and
a motivational argument for basing our approach on this formalism hinges, on a nut shell,  on the theoretical observations that since physically meaningful quantities should be independent of the choice of a gauge, the concepts of gauge potentials or connections had to be incorporated into the formulation of Action Densities for describing our perception of Nature. This then has led naturally to the formalism of fiber bundles to describe the basic forces of nature and the mathematical physics for dealing with Gauge Theory and Variational Principles in Field Theory. Now, a bundle $P(M,F,\tau)$ consists of a topological space $P$, a base $M$, a typical fiber $F$ and a continuous surjection $\tau:P\to M$, where in semi-classical physics $M$ is the space-time continuum with a Hausdorff topology. Moreover, it
can be shown that a vector bundle over $M$ can be described purely in terms of concepts pertinent to the commutative ${\mathcal C}^\star$-algebra  $C(M)$ (see {\it e.g.}\cite{lands}). Furthermore, by the Gel'fand-Naimark Theorem \cite{gel-naim}: ``To every commutative  ${\mathcal C}^\star$-algebra with unit there corresponds a Hausdorff space, which implies a complete duality between the category of locally compact Hausdorff spaces and the category of commutative
${\mathcal C}^\star$-algebras $C(M)$ and $^\star$-homomorphisms. However, at distances of the order of the Planck length, where the Principle of Uncertainty and the Principle of Equivalence become equally important and noncommutativity dominates the dynamics of the system, one needs to generalize the notion of a Hilbert bundle in such a way that the commutative ${\mathcal C}^\star$-algebra $C(M)$ is replaced by an arbitrary ${\mathcal C}^\star$-algebra $\frak A$, and the dual notion of a Hausdorff topological space $M$ be replaced by the space of all unitary classes of irreducible representations of $\frak A$ (\cite{connes2}, \cite{rieffel1},\cite{rieffel2},\cite{rieffel3}).

On the basis of the previous remarks and in order to implement this ideas so as to provide the possibility of calculation for observable quantities in physical models,  the material in this paper has been structured as follows:  In
Section II we introduce a projective unitary realization of the generators of the twisted discrete translation group $\mathcal C^\star$-algebra $\frak A$ of bounded operators with unit, $^\star$-homomorphic to the Heisenberg-Weyl group of deformed quantization. Thus the noncommutative lattices, generated from the primitive spectrum of $\frak A $, are the structure spaces of the $T_{0}$ Jacobson topology  and the noncommutative analogue of the Hausdorff topology of the space $M$ of the Gel'fand - Naimark theorem.
In Section III we go on to use the homomorphism obtained in the previous section and the Gel`fand-
Naimark-Segal construction to derive the kinematic Hilbert space on which the bounded operators in $\frak A$ will act.
In addition, the functions resulting from the Pontryagin duality on this Hilbert vector space yield a complete set of functions which satisfy the same orthogonality and summation completeness relations as the algebra of almost periodic functions \cite{thie}.
Section IV  begins by considering the ADM reduced classical action
of the anisotropic Bianchi I model cosmology coupled to a massless
scalar to assume the part of an inner time. We then quantize the
system following Dirac's procedure after expressing the observables
of the system in terms of the $\mathcal C^\star$-algebra of
Hermitized bounded operators previously introduced. Using then the
Hamiltonian constraints of the system and applying well documented
techniques such as the ones summarized and cited in the text, we
derive the physical states of the system from the kinematical states
constructed in Sec.III. In Section V the so far inherently discrete
system of equations is converted to the continuum by making use of
the Feynman Path Integral construction for quantization. It should
be noted, however, that the symbol of noncommutativity appears in
various terms of the action and acquires different levels of
relevance for the different possible stages of evolution of the
system, as shown in the later sections. This analysis is in fact
carried out extensively in Sections VI and VII, after deriving the
equations of motion by applying the method of stationary phase to
the action derived in Sec.V. In Section VII, in particular, we
consider several scenarios for the system evolution which evidence
clearly that noncommutativity, in the form that we have introduced
here, not only prevents the singularities that occur in the
Classical and Wheeler-DeWitt quantization approach to the Bianchi
Cosmology, but it also provides the driving force which, under
appropriate boundary conditions, allows the system to leave from a
stage of oscillatory evolution within Planck length scales, to
stages of regions where noncommutativity becomes negligible and the
universe growth is monotonical. In Sec. VIII we summarize what we
consider are the main results of this work and possible future lines
of research that would extend it.

\section{Twisted Discrete Translation Group $C^\star$-algebra and Deformation Quantization}
Let us now consider  \cite{bedos1}, \cite{paker}, \cite{bedos2}
the twisted (unital, discrete) ${\mathcal C}^\ast$-dynamical system $\Sigma=({\mathcal A}, G, \alpha,\sigma)$ where the algebra $\mathcal A$
can be related  by means of a *-homomorphism to the $C^\ast$-algebra ${\frak A}\subset {\mathcal B}(\mathcal H)$ of bounded operators with unit, acting on a Hilbert space $\mathcal H$. For this purpose and
as a starting point of our analysis we observe that, since the base topological $M$
space in
Classical Bianchi I Cosmology is an ${\mathbb R}^{3}$, for which translations are isometries, whereas physical space at the Noncommutative Geometry level is described as a sort of a subjacent discrete
noncommutative cellular structure (posets), we let ${\mathcal A}$ be the algebra of the noncommutative extended Heisenberg-Weyl group \cite{rm},  $G$ be the discrete topological group of translations in ${\mathbb R}^{3}$, $(\alpha, \sigma)$ the twisted action of $G$ on ${\mathcal A}$, with $\alpha$ denoting the map $\alpha:G\to \text {Aut} ({\mathcal A})$ and $\sigma:G\times G\to {\mathcal T}({\mathcal A})$
 is a normalized 2-cocycle on $G$ with values in the multiplicative group ${\mathcal T}$ of all complex numbers of unit modules, such that
\begin{align}\label{cocycle}
\sigma({\bf x}_1, {\bf x}_2)\sigma({\bf x}_1 +{\bf x}_2, {\bf x}_3)&=\sigma({\bf x}_2, {\bf x}_3)\sigma({\bf x}_1 , {\bf x}_2 + {\bf x}_3), \;\;\;\;\; {\bf x}_1 ,  {\bf x}_2,  {\bf x}_3\in G\nonumber\\
\sigma({\bf x}, {\bf 0})=\sigma({\bf 0}, {\bf x})&=1.
\end{align}
In the above we have identified the discrete Abelian group of translations $G$ with the vector space ${\bf T}_3$, associated with  ${\mathbb R}^{3}$ as an affine space with a discrete topology and with coset decomposition
\be\label{transl}
{\bf T}_3 =\sum^{\infty}_{j_1 ,j_2 , j_3=-\infty} (\mu_i j_i) {\hat e}_{i}, \;\;\;\;\; j_i \in {\mathbb Z},
\ee
where the ${\hat e}_{i}$ are the basic translations in ${\mathbb R}^{3}$,  the vectors $ {\bf x}_{(l)} = \sum_{i=1}^3(\mu_i j_{(l)i}){\hat e}_{i}\in{\bf T}_3 $
are elements of ${\mathbb R}^{3}$ as a group and the set $\Gamma:\{\mu_i j_{(l)i} \}$ form a 3-dimensional cell. We then have

{\bf Definition II.1}. {\it A left $\sigma({\bf x}_1, {\bf x}_2)$-projective unitary representation $\hat U$ of $G$ on a (non-zero) Hilbert space $\mathcal H$ is a map
 from the group $G$ into the group ${\mathcal U}(\mathcal H)$ of unitaries on
$\mathcal H$ such that

\be\label{comm2}
 U ({\bf x}_1)  U ({\bf x}_2)=\sigma({\bf x}_1, {\bf x}_2) U({\bf x}_1+{\bf x}_2) .
\ee}
Taking in particular
\be\label{sigmateta}
{\mathcal U}(\mathcal H)\ni\sigma_{\theta}({\bf x}_1, {\bf x}_2) :=\sigma({\bf x}_1, {\bf x}_2)=e^{-i\pi {\bf x}^{T}_1 R\, {\bf x}_2}= e^{-i\pi\boldsymbol\theta\cdot ({\bold x_1}\times {\bold x_2})},
\ee
where $R$ is the anti-symmetric matrix\\
\be\label{matrix}
R=
\begin{pmatrix}
0& \theta_3 & -\theta_2\\
-\theta_3 & 0& \theta_1\\
\theta_2 & -\theta_1 & 0
\end{pmatrix},
\ee
where the $\theta_i$ have been assumed to be Poincar\'e invariant, as shown in \cite{chaich}, when considering a deformation of the universal enveloping Hopf algebra $\mathcal U(P)$ of the Poincar\'e algebra $\mathcal P$ by means of a Drinfeld twist \cite{drinf1}.\\
{\bf Definition II.2}. {\it A left projective regular unitary realization of the algebra (\ref{comm2}) and (\ref{sigmateta})  on $l^2(G)$ can be defined as}
\be\label{Lambda}
\langle {\bf x}|\hat U_i  |\xi\rangle:= e^{-2\pi i\varepsilon_i x_i} \langle {\bf x} -\frac{1}{2}\varepsilon_i\hat e_i \times\boldsymbol\theta|\xi\rangle = e^{-2\pi i\varepsilon_i x_i}\xi({\bf x} -\frac{1}{2}\varepsilon_i\hat e_i \times\boldsymbol\theta); \;\;\;\;
\xi({\bf x}) \in {\mathcal H}.
\ee
Identifying $\bf x$ with the corresponding function on ${\bf T}_3$ which is one at $\bf x$ and zero otherwise, {\it i.e.} if we let this function be  $\delta_{\bf x}\in l^2({\bf T}_3)$ (the delta function at ${\bf x}$) then
it readily follows that
\be\label{Lambda1}
\hat U_i \delta_{\bf x}:= e^{-2\pi i\varepsilon_i x_i}\delta_{(\frac{1}{2}\varepsilon_i\hat e_i \times\boldsymbol\theta +\bf x)},
\ee
and
\be\label{Lambda2}
\hat U_i |{\bf x} \rangle = e^{-2\pi i\varepsilon_i x_i} |{\bf x} + \frac{1}{2}\varepsilon_i\hat e_i \times\boldsymbol\theta\rangle.
\ee
Thus the unitary $\hat U_i$ translates the vector $\bf x$ in a direction perpendicular to $\hat e_i$ by the amount $\frac{1}{2}\varepsilon_i\boldsymbol\theta$. It is now fairly straightforward to show, by successive applications of (\ref{Lambda}), that
\be\label{unitaries1}
\hat U_i \hat U_j =e^{-i\pi\varepsilon_i\varepsilon_j {\boldsymbol\theta}\cdot ({\hat e}_{i}\times {\hat e}_{j})} \hat U_{i+j},
\ee
and interchanging indices and substituting back the result into (\ref{unitaries1}) we arrive at
\be\label{unitaries2}
\hat U_i \hat U_j =e^{-2i\pi\varepsilon_i\varepsilon_j {\boldsymbol\theta}\cdot ({\hat e}_{i}\times {\hat e}_{j})}\hat U_j \hat U_i .\
\ee

Since the parameter of noncommutativity actually has units of length square the quantities $\varepsilon_i$ must have units of $\text {length}^{-1}$ and $\varepsilon_i\hat e_i \times\boldsymbol\theta$
are thus basic vectors in the directions perpendicular to the $\hat e_i$ which determine the fundamental lengths of the lattice.\\
Extending now the above algebra with the generators $\hat V_l :=\hat V({\mu_l \hat e}_l)$ such that
\be\label{vmult}
\hat V_l |{\bf x} \rangle =|{\bf x} +\mu_l \hat e_l\rangle,
\ee
so we find that $\hat V_l$ also acts on the kets $|{\bf x}\rangle \in{\mathcal H}$ as a translation operator on the vector $\bf x$ in the direction of $\hat e_l $ by an amount $\mu_l$.
It also follows from (\ref{vmult}) that
\be\label{v2}
\hat V_i \hat V_l= \hat V_l \hat V_i ,
\ee
and commuting with $\hat U_i$ as given in (\ref{Lambda2}), we arrive at
\be\label{v1}
\hat U_i \hat V_l  =e^{-2\pi i\varepsilon_i \mu_l ({\hat e}_{i} \cdot {\hat e}_{l} ) } \hat V_l \hat U_i =e^{-2\pi i\varepsilon_i \mu_l\delta_{il}}\hat V_l \hat U_i.
\ee
This is indeed a *-homomorphism between the ${\mathcal C}^\ast$-algebra ${\frak A}\subset {\mathcal B}(\mathcal H)$ of operators generated by the unitaries $\hat U_i$'s and $\hat V_l $'s  and
the extended noncommutative Heisenberg-Weyl algebra $\mathcal A$
 of the  ${\mathcal C}^\star$-dynamical system discussed before. \
Note also that the quantities $\mu_l$ and $\varepsilon_i$ introduced in the above relations
strictly appear so far as independent parameters of the action of the discrete subgroups of the  twisted (extended noncommutative) Heisenberg-Weyl group.
This would however imply two different simultaneous noncommutative lattices
generated by the unitaries $\hat U_i$'s and $\hat V_l $'s. Clearly in order to avoid this the
$\mu_l$ and $\hat e_l \cdot(\varepsilon_i\hat e_i \times\boldsymbol\theta)$
must be related. We shall show later on that this relation appears naturally
when constructing the Hilbert space on which these operators act.

We also find it important to point out here that, although the
expressions (\ref{unitaries1}) and (\ref{unitaries2}) for the
subalgebra of the $\hat U_i$ appear to be the same as that used to
describe the quantum torus ({\it cf. e.g.} \cite{bedos3}), the
realization  (\ref{Lambda}) (or (\ref{Lambda2})) introduced here has
quite different implications. Indeed, as mentioned in the paper
cited above, in the quantum torus formulation the $\hat U_i$ act as
Laplacian operators that translate on momentum space, and thus are
appropriate to describe noncommutativity in momentum space \cite{dmt}. On the
other hand the realization of the $\hat U_i$ and $\hat V_l$
unitaries in (\ref{Lambda2}) and (\ref{vmult}) is geared to generate
a Hilbert space by sequential translations, effected by the
noncommutation matrix factor, on a cyclic vector. Thus in this case
the noncommutativity is associated with the dynamical configuration
variables of our formulation. The strong
repercussions for our developments of this choice of realization is
evidenced in the analysis  presented in the last sections of this
work.

\section{GNS-Construction of the Kinematic Hilbert Space}

Let us now use this homomorphism to derive explicit forms for the elements of the Hilbert space $\mathcal H$ on which the operators in $\frak A$ act by applying the Gel`fand -Naimark-Segal (GNS) construction \cite{woro},\cite{landi}. To this end first note
that for any state functional $\phi$ we have that $\forall\; a \in {\mathcal A}$ $\exists \; \phi$ such that $\phi(a^* \star a)=1$. Moreover, since any element $a$ in the subjacent algebra ${\mathcal A}$ is unitary, we have that this equality is always true here which, in turn, implies that the left ideal  ${\mathcal I} =\{ a \in {\mathcal A}\; |\phi(a^{\ast} \star a)=0\}$ in ${\mathcal A}$ is empty, so that the quotient space
${\mathcal N}_\phi ={\mathcal A}/{\mathcal I}_{\phi} \equiv {\mathcal A} \Rightarrow\phi$  is faithful. Thus, by
the GNS construction, we have a pre-Hilbert space with a non-degenerate product defined by
 \be\label{prehil}
{\mathcal A}\times {\mathcal A}\to \mathbb C, \;\;\;\;\;\; \langle a, b \rangle\mapsto \phi(a^{\ast}\star b),
\ee
and where ${\mathcal H}_\phi$ is the completion of $\mathcal A$ in this norm.
Note that the $\star$-homomorphism $\pi_\phi : {\mathcal A}\to {\mathcal B}(\mathcal H_\phi)$, defines a
representation $({\mathcal A}, {\mathcal H}_\phi)$ of the $C^\star$-algebra ${\mathcal A}$ by associating
to an element $a\in {\mathcal A} $ an operator $\pi_\phi(a)) \in {\frak A}\subset{\mathcal B}(\mathcal H)$ by
\be\label{pirep}
\pi_\phi(a) b =a \star b,
\ee
which is a well defined bounded linear operator in ${\mathcal H}_\phi$. Indeed, from the above definition it follows that
\be\
\pi_\phi(a_1) \pi_\phi(a_2) (b)=a_{1}\star a_{2}\star b =\pi_{\phi} (a_1 \star a_2)b,
\ee
which shows that (\ref{pirep}) is in fact a representation. Note also that in this construction the $C^\star$-algebra
is itself a Hilbert $\mathcal A$-module.

Now, in order to generate the elements of the Hilbert space we start with a distinguished vector $\xi_\phi$ which
is cyclic for $\pi_\phi$, {\it i.e.} such that $\{\pi(a)\xi_\phi| a \in {\mathcal A}\}$
is dense in ${\mathcal H}_\phi$. Since $\mathcal A$ is unital we can chose $\xi_\phi:= \langle {\bf x}=0 |\xi_\phi\rangle =\xi_\phi (0,0,0)=I$, which is clearly cyclic
provided the parameters $\varepsilon_i$ and $\mu_l$, generated by the operators $\pi_\phi (a)= \hat U_{i} , \;\hat V_l$ $\in {\mathcal B}(\mathcal H_\phi)$, according to (\ref{Lambda2}) and (\ref{vmult}) and which translate in directions perpendicular to each other, are appropriately related in order that the set of elements generated by the action of the $\pi_\phi (a)$ on $\xi_\phi$ is indeed dense in ${\mathcal H}_\phi$. It is not difficult to show
that such a consistency can be achieved by setting

\begin{align}\label{rel3}
\mu_1 = & \frac{n_1}{2} \varepsilon_2 \theta_3\nonumber\\
\mu_2 = & \frac{n_2}{2} \varepsilon_1 \theta_3  \\
\mu_3 = & \frac{n_3}{2} \varepsilon_1 \theta_2,  \nonumber
\end{align}
where, as we shall show later on in Section VII, the magnitudes
  $ n_i \in \mathbb N^{+}$  and $\bar\varepsilon_{i}$  are scale factors of the $\mu_i$'s and $\varepsilon_{i} $'s
determined by the relative relevance of the noncommutative tensor
symbol in the different stages of evolution of the dynamical system
that we shall consider later on. In fact, we can consider the $\mu_i$'s and $\varepsilon_{i} $'s  as introduced in the formalism to effectively
represent a family of continuous projections $\pi^{m,n}$ acting on a family of topological spaces $Y^{n}$ such that
\be\label{pro}
\pi^{m,n}: Y^{m}\to  Y^{n}, \quad n\leq m.
\ee
Hence the manifold $M$ with Hausdorff topology $(Y^{\infty})$ can be recovered as the limiting procedure of the inverse of such a sequence of projectors \cite{bim}.
Moreover, in the limit $\varepsilon_i \to 0$ it readily follows that (\ref {Lambda2}) becomes multiplicative and the $\mu_{l}$ decouple from (\ref{rel3}) and
(\ref{rel4}), so our twisted Heisenberg-Weyl algebra reduces to that in \cite{ash1} and the commutative lattices generated by the primitive spectrum of this algebra are now structure spaces of a $T_1 $ topology where, as we shall show later on in Sec.VI, the elementary length of the cell induced by the $\mu_l$'s is of
${\mathcal O}(\lambda_P)$. Taking the further limit $\mu_l \to 0$ will then result in the classical Heisenberg-Weyl algebra and a Hausdorff or $T_2$-space.\\

Note also that in some sense the relations (\ref{rel3}) are an equivalent of the improved dynamics introduced in
\cite{ash2}, which in  our case appear directly from the consistency
required by the translations generated by the noncommutativity. From
(\ref{rel3}), (\ref{Lambda2}), and (\ref{vmult}) we also get
\begin{align}\label{rel4}
\varepsilon_2 \theta_3 =&\varepsilon_3 \theta_2 \nonumber\\
\varepsilon_1 \theta_3=&\varepsilon_3 \theta_1 \\
\varepsilon_1 \theta_2=&\varepsilon_2 \theta_1 \nonumber.
 \end{align}
Consequently, it follows from the above relations that
 the subset $\{\pi(\hat V_i) \xi_{\phi}\}$ will be by  itself dense in $\mathcal H_\phi$ and, by virtue of (\ref{pirep}) and (\ref{prehil}) (and the GNS Theorem),
we have that given a vector-state functional $\phi$ on $\{V_l\}\subset \mathcal A$ there is a $\star$-representation  with a distinguished cyclic vector $\xi_\phi \in \mathcal H_\phi$ with the property
\be\label{vstate}
\langle\xi_{\phi} , \pi_\phi (V_l)\xi_{\phi}\rangle =\langle {I}, V_l\rangle =\phi(V_l).
\ee
Recall now that (\ref{vmult}) implies that
\be\label{Lambda3}
\langle {\bf x}_1 ={\bf 0}|\hat V_{l} |\xi_\phi\rangle =\xi_\phi ({\bf0} +\mu_l \hat e_l)= \xi_\phi (\mu_l \hat e_l),
\ee
so, if via the algebra *-homomorphism we associate to the element $V_{l} \in {\mathcal A}$ the operator \\
$\pi_\phi (V_{l}) = \hat V(-\mu_l \hat e_l))$, then combining (\ref{vstate}) with(\ref{Lambda3}) allows us to identify $\phi(V_l)$ with the character of the discrete translation group, so that
\be\label{char}
\xi^k_\phi({\bf x}_n) = e^{2\pi i \sum_{l=1}^3 {\mu_l}({k_l} j_{(n)l}) },\;\;\;\;\;j_{(n)l}\in\mathbb Z
\ee
where ${\bf k} \in {\mathbb R}^3$, and $\mu_l$ are quantities whose magnitudes determine the size of the fundamental noncommutative lattice cell. Observe also that, since $\mathcal I$
is empty, the representation $(\mathcal H_\phi, \xi_\phi)$ is irreducible.

The functions $\xi^k_\phi({\bf x})$ in (\ref {char}) are a one-dimensional irreducible regular representation of the operator group $\bar D^{\bf k} ({\bf x})$
 of the discrete Abelian group of translations. That is
\be\label{re}
\bar D^{\bf k} ({\bf x}_n) = e^{2\pi i \sum_l{\mu_l}({k_l} j_{(n)l}) },
\ee
and satisfies the relations of orthogonality and Poisson summation completeness  \cite{L&K}
\begin{align}\label{ortho}
\int_{-1/2\mu_{l}}^{1/2\mu_{l}}\mu_l dk_l \;{\bar D}^{ k_l} (j_{(1)l} )  D^{ k_l} (j_{(2)l})& =\delta_{j_{(1)l}{j_{(2)l}}} ,\;\;\;l=1,2,3\nonumber\\
\sum^{\infty}_{j_i-\infty} {\bar D}^{ k_i} \left( j_i \right )  D^{{ k_i}^{\prime}}\left( j_i  \right)& =\sum_{m_i=-\infty}^{\infty} \delta(\mu_i  k_i -\mu_i k_i^\prime +m_i ),
\end{align}
respectively, after noting that the left hand side of the second equation above is a periodic generalized function with period one \cite{light}. Observing that since the representations (\ref{re}) of the translation group are invariant under the reciprocal group, the range of fundamental domain of the components of the vector
parameter $\bf k$ is $- 1/2\mu_i \leq k_i \leq 1/2\mu_i $.\\
Also, making use of the completeness of the ket space $\{|{\bf k} \rangle\}$ we can write
\be\label{val}
\bar D^{k_l} (j_{(n)l})=e^{2 i\pi j_{(n)l}\mu_l k_l}: = \langle \mu_lj_{(n)l}| k_l\rangle=\langle x_{(n)l}|k_l\rangle,
\ee
with
\be\label{intval}
\prod_{l=1}^{3} \int^{\frac{1}{2\mu_l}}_{-\frac{1}{2\mu_l}}\mu_l dk_l \langle x_{(n)l}|k_l\rangle\langle k_l|x_{(n')l}\rangle=:
\langle\bold x_{(n)}|\bold x_{(n')}\rangle = \delta_{\bold x_{(n)},\bold x_{(n')} }.
\ee
Furthermore, by the Pontryagin duality
theorem, the dual of a discrete Abelian group is a compact Abelian group, so by Fourier analysis we can write (for a fixed index $i$)
\be\label{FT1}
{\hat f}( k_i)= \sum_{j_{(l)i}=-\infty} ^{\infty} f(j_{(l)i})\; e^{\mu_i j_{(l)i} (2i\pi k_i) }, \;\;\;\;  - 1/2\mu_i \leq k_i \leq 1/2\mu_i, \;\;\;\;i=1,2,3,
\ee
and
\be\label{FT2}
f( j_{(l)i})=\int_{-1/2\mu_i}^{1/2\mu_i} dk_i \; {\hat f}( k_i) \;e^{-k_i (2i\pi\mu_i j_{(l)i}) }.
\ee

Denote by $\Gamma =\{e^{k_i (2i\pi\mu_i j_{(l)i}) }\}$ the compact Abelian group of continuous characters dual to the twisted discrete translation group $G$, and let $\bar G$ denote the Abelian compact group of all characters, continuous or not, of $G$. Then $\Gamma $ is a continuous isomorphism of  $G$ onto a dense subgroup $\beta(G)$ of  $\bar G$. Thus, since the generators $e^{(2i\pi { k_i})}$ of the basis of mono-parametric subgroups in (\ref {FT1}) are isomorphic to the circle group ${\mathcal T}$ we have that the ${\hat f}( k_i)$ in (\ref{FT1}) can be regarded as elements of the dense subgroup of the Bohr compactification of the  twisted discrete translation group onto the quantum 3-torus =$\bar G$. \\


In particular, setting $x_{(l)i}:=\mu_{i} j_{(l)i}$ we see that the function $ e^{2i\pi x_{(l)i} k_i }$ is continuous and periodic in $ k_i $,
thus the polynomial function $\sum_{l=1}^N f( x_{(l)i})\; e^{-2i\pi x_{(l)i}  k_i }$ is an almost periodic function in
the sense of Bohr ({\it cf.} \cite{anzai} \cite{EDM}). Furthermore
if the latter function converges uniformly to the series $\sum_{l=1}^{\infty} f( x_{(l)i})\; e^{2i\pi x_{(l)i}  k_i }$
when $N\to \infty$, then the limit function is also almost periodic.
Next note that if we now introduce the reciprocal group of the discrete group of translations on the reciprocal lattice
\be
L^R:= \{b^R = b_i/\mu_i,\;\;\; b_i \in {\mathbb Z} \},
\ee
 it follows immediately from (\ref{FT1}) that
\be\label{period}
{\hat f}( k_i) ={\hat f}( k_i + b_i/\mu_i),
\ee
which confirms the statement below equation (\ref{ortho}) regarding the fundamental domain of $k_i$.
In summary, we have seen that the space-space noncommutativity of the Heisenberg algebra can be expressed by a realization of the associated Heisenberg-Weyl group by a $C^\ast$-algebra ${\frak A}\subset {\mathcal B}(\mathcal H)$ of bounded unitary operators with unit, acting on a non-separable Hilbert space where an orthonormal basis is the set of almost periodic functions :
\be\label{basis}
\{\xi^k_{\phi} ({\bold x}_{(l)})=\bar D^{\bold k} ({\bold x}_{(l)}) =e^{2i\pi \bold x_{(l)}{\bold\cdot}\bold k}\},
\ee
given by the characters in (\ref{char}).

\section{Quantum Cosmology for the anisotropic Bianchi I model}

As it is well known the classical action function, after ADM reduction to canonical form, for a Bianchi I cosmology describing a gravitational field, with space-time metric
\be\label{metric}
g_{\mu\nu}=
\begin{pmatrix}
-N^2 (t)& 0& 0& 0\\
0& a_1^2 (t)&0&0\\
0&0&a_2^2 (t)&0\\
0&0&0&a_3^2 (t)
\end{pmatrix},
\ee
minimally coupled to a massless scalar field $\varphi (t)$ independent of the spatial coordinates,
is given by
\begin{align}\label{taction1}
S_{grav}+ S_\varphi  &= \left(\frac{c^3}{G}\right)\int\left(\pi^{ij} \dot g_{ij} -\frac{N(t)}{\sqrt {^3 g}} \left[-\frac{1}{2}(\pi^k_k)^2 + \pi^{ij}\pi_{ij} \right]\right) d^4 x\nonumber\\
& + \hbar\int d^4 x \left(p_{\varphi }\dot\varphi -\frac{1}{2}\frac{N}{\sqrt{^3 g}}p^2_{\varphi }\right) ,
\end{align}
where ({\it cf.} Chapter 21 of \cite{misner}) the tensor densities
$\pi^{ij}$ are the canonical momenta conjugate to the metric
components $g_{ij}=a^2_i(t)$ (the square of the Universe radii),
$N(t)$ is the lapse function and $p_\varphi $ is the canonical
momentum conjugate to $\varphi $, with $p_\varphi $ being in units
of length and $\varphi $ in units of inverse of length . Moreover,
writing the kinematic term  in (\ref{taction1}) as $\pi^{ij} \dot
g_{ij}= 2\pi^{ii}a_i \dot a_i$ and making the definition
$2\pi^{ii}a_i:=\pi^i$ we can re-express the gravitational action in
(\ref{taction1}) in the form \be\label{taction2} S_{grav} =
\frac{1}{2}\left(\frac{c^3}{G}\right)\int\left(\pi^{i} \dot a_{i}
-\frac{N(t)}{2\sqrt {^3 g}}
\left[-\frac{1}{2}\left(\sum_{i=1}^3\pi^i a_i\right)^2 +
\sum_{i=1}^3(\pi^i{a_i}^2 \pi^{i}) \right]\right) d^4 x,
\ee
or, observing next from equation (21.91) in \cite{misner} that $\pi^{ij}$ is unitless and therefore that
$\pi^i$ has units of length, we can define a new quantity $p^i:=\frac{c^3}{G\hbar}\pi^i$,
which has units of inverse of length, so (\ref{taction2}) can be written as
\be\label{taction3}
S_{grav}= \frac{1}{2}\hbar\int\left(p^{i} \dot a_{i} -\frac{N(t)}{2\sqrt {^3 g}}\left(\frac{G\hbar}{c^3}\right)
\left[-\frac{1}{2}\left(\sum_{i=1}^3 p^i a_i\right)^2 + \sum_{i=1}^3(p^i
{a_i}^2 p^{i}) \right]\right) d^4 x.
\ee In addition, the scalar field action can be re-expressed as:
\be\label{taction4} S_\varphi =\hbar\int d^4 x \left(p_{\varphi
}\dot\varphi - \frac{1}{2}\frac{N}{\sqrt{^3
g}}\left(\frac{G\hbar}{c^3}\right)\left(\frac{c^3}{G\hbar}\right)p^2_{\varphi
}\right), \ee and defining \be\label{taction5} p_{\phi} :=
\left(\frac{c^3}{G\hbar}\right)^{\frac{1}{2}} p_{\varphi },\;\;\;\;
\text{and} \;\;\;\;
\dot\phi:=\left(\frac{G\hbar}{c^3}\right)^{\frac{1}{2}}\dot\varphi ,
\ee where both $p_{\phi}$ and $\dot\phi$ are unitless, we arrive at
\be\label{taction6} S_\phi=\hbar\int d^4 x
\left(p_{\phi}\dot\phi-\frac{1}{2}\frac{N}{\sqrt{^3
g}}\left(\frac{G\hbar}{c^3}\right)p^2_{\phi}\right). \ee
Consequently the total classical Hamiltonian constraint is
\cite{chiou}, \cite{salis}: \be\label{ccont1} C_{\text grav} +
C_\phi=\frac{N(t)}{2\sqrt {^3
g}}\left(\frac{G\hbar}{c^3}\right)\left[\left(-\frac{1}{2}\left(\sum_{i=1}^3
p^i a_i\right)^2 + \sum_{i=1}^3(p^i{a_i}^2 p^i)\right)
+\frac{1}{2}p_\phi^2 \right] =0. \ee

If we choose the lapse function to be $N(t)(4 (^3
g))^{-\frac{1}{2}}=\left(\frac{c^3}{G\hbar}\right) $
and assume for simplicity the following ordering for the quantum Hamiltonian constraint operator, we therefore have:
\be\label{qconst1}
\hat C=\hat C_{\text grav} + \hat C_\phi
= \frac{1}{2}\left( -\sum_{i\neq j} \hat p^i \hat p^j \hat a_i \hat a_j + \sum_{i}\hat p^{i}  \hat a^2_i \hat p^{i}\right)
+ \frac{1}{2}{\hat p_\phi^2}=\hat 0.
\ee

Now, since the action of the $\hat p^i$ and $\hat a_i$ operators on our Hilbert space
basis of kets is to be derived from the unitary operator representations discussed in the previous
section and whose action on the Hilbert space is displayed in equations (\ref{Lambda2}) and (\ref{vmult}). For this purpose
it is important to notice that the Hilbert space is constructed from the noncommutative group of operators $\frak A$. Moreover,
due to the noncommutativity, the elements of this group are not exponentials of self adjoint operators. To construct the observables
$\hat a_i $ we thus take
\be\label{repa}
\hat a_i : = -\frac{\hat U_i - \hat U^\dagger_i}{2 i\varepsilon_i},
\ee
 so that
\be\label{op7}
\hat a_{i} |{\bf x}_{(n)}\rangle = -\frac{1}{2 i\varepsilon_i}\left( e^{-2i\pi\varepsilon_i x_i} |{\bf x}_{(n)}+\frac{1}{2}\varepsilon_{i} {\hat e}_{i}\times \boldsymbol\theta\rangle -
e^{2i\pi \varepsilon_i x_i} |{\bf x}_{(n)}-\frac{1}{2}\varepsilon_{i} {\hat e}_{i}\times \boldsymbol\theta\rangle\right),
\ee
and
\be\label{op6}
\hat p^l  :=  \left(\frac{V_l(\mu_l)-V^\dagger_l(\mu_l)}{2 i \mu_l}\right).
\ee
so that
\be\label{op8}
\hat p^l |{\bf x}\rangle = \frac{1}{2 i \mu_l}\left(|{\bf x} + \mu_l \hat e_l\rangle -|{\bf x} - \mu_l \hat e_l\rangle\right).
\ee

That (\ref{repa}) reproduces the uncertainty principle for mean-square-deviations
of the distributions $\langle\Psi| \hat a_i |\Psi \rangle$ and the noncommutative
algebra  of the $\hat a_{i}$ for the discrete case, can be seen
by substituting (\ref{repa}) in the commutator $[\hat a_{i}, \hat a_{l}]$
and making use of (\ref{Lambda2}) and (\ref{unitaries1}). We then find that
\begin{align}\label{uncert5}
\langle {\bf j'}| [\hat a_i , \hat a_l ] |{\bf j}\rangle = &\left(\frac{2i}{\varepsilon_i \varepsilon_l}\right)\sin(\pi\varepsilon_i \varepsilon_l \boldsymbol\theta\cdot(\hat e_i \times\hat e_l) )\prod_{m=1}^{3}\int_{-\frac{1}{2}}^{\frac{1}{2}}
d\bar k_m\;e^{2\pi i \bar{\bf k}\cdot ({\bf j' -{\bf j}})} \cos\left( 2\pi \varepsilon_i \mu_i [j_i + (\frac{1}{2\mu_i}){\bf k}\cdot (\hat e_i \times \boldsymbol\theta)]\right)\nonumber\\
&\times \cos\left( 2\pi \varepsilon_l \mu_l [j_l + (\frac{1}{2\mu_l}){\bf k}\cdot (\hat e_l \times \boldsymbol\theta)]\right) \;\;\;\;\;\;\text {where}\;\; \bar k_m := \mu_m k_m,
\end{align}
from where it can be inferred that the quantity
\be\label{uncert6}
\left(\frac{2}{\varepsilon_i \varepsilon_l}\right)\sin(\pi\varepsilon_i \varepsilon_l \boldsymbol\theta\cdot(\hat e_i \times\hat e_l) ) \cos\left( 2\pi \varepsilon_i \mu_i [j_i + (\frac{1}{2\mu_i}){\bf k}\cdot (\hat e_i \times \boldsymbol\theta)]\right)
\times \cos\left( 2\pi \varepsilon_l \mu_l [j_l + (\frac{1}{2\mu_l}){\bf k}\cdot (\hat e_l \times \boldsymbol\theta)]\right)
\ee
is the symbol of the action of the operator commutator on the spectral representation of the product $\langle {\bf j'} |{\bf j}\rangle$.
In the limit $\varepsilon_i \varepsilon_l \boldsymbol\theta\cdot(\hat e_i \times\hat e_l)<<1$ (since by (\ref{rel3}) and (\ref{rel4}) also implies $\varepsilon_i \mu_i <<1$ ) , the above symbol of $[\hat a_{i}, \hat a_{l}]$ is $2\pi  \boldsymbol\theta\cdot(\hat e_i \times\hat e_l) $.

The expressions (\ref{op7}), (\ref{op8}),  are to be substituted into (\ref{qconst1}) in order to derive the action of the constraint operator on the Hilbert vectors $|{\bf x}_{(n)}\rangle$.\

To make a detailed connection with other formulations
we use the Feynman phase space path integral procedures
considered in \cite{ash1}. The general idea of the group averaging procedure (see {\it e.g.} \cite{marolf}) is that
the physical state $|\Psi_{phys}\rangle\in {\mathcal H}_{phys}$, which is a solution of the constraint equation, is derived by averaging the action of the unitary monoparametric Abelian group $\exp(i\alpha \hat C),\;\; \alpha\in\mathbb R $,
on a state $|\Psi_{kin}\rangle$ in an auxiliary kinematic Hilbert space $\mathcal H_{kin}$ dense in ${\mathcal H}_{phys}$. Thus
\be\label{ga}
|\Psi_{phys}\rangle = \int_{-\infty}^\infty d\alpha\; \exp(i\alpha \hat C)|\Psi_{kin}\rangle.
\ee
Heuristically (\ref{ga}) can be justified as a refined algebraic quantization by observing that the integrand can be viewed as a Fourier Dirac delta representation:

\be\label{ga2}
\int_{-\infty}^\infty d\alpha\; \exp(i\alpha \hat C) \sim \delta (\hat C),
\ee
and that by acting on (\ref{ga}) with $U(\beta) = \exp(i\beta\hat C)$ we have

\begin{align}
U(\beta)|\Psi_{phys}\rangle & =  \exp(i\beta \hat C)\delta(\hat C)|\Psi_{kin}\rangle=\delta(\hat C)|\Psi_{kin}\rangle\nonumber\\
&=\int_{-\infty}^\infty d\alpha\; \exp[i(\alpha + \beta) \hat C]|\Psi_{kin}\rangle=\int_{-\infty}^\infty d\alpha'\; \exp(i\alpha' \hat C)|\Psi_{kin}\rangle=|\Psi_{phys}\rangle,
\end{align}
therefore the unitaries $U(\beta)\;\forall\beta$ act trivially on the physical states defined as in (\ref{ga}), consistent with Dirac's requirement that physical states be annihilated by the constraints.
however, the physical state defined by (\ref{ga}) is not normalizable. Hence, in order to eliminate one of the deltas in the inner product, this is defined according to
\be\label{ga3}
(\Phi_{phys}|\Psi_{phys}):= \int_{-\infty}^\infty d\alpha\; \langle \Phi_{kin}|\exp(i\alpha \hat C)|\Psi_{kin}\rangle.
\ee
Clearly this definition of the inner product has the advantage that it remains the same for any two other physical states of the form $|\Phi'_{phys}\rangle =\exp(iu\hat C)|\Phi_{phys}\rangle $.\

Now, an orthonormal basis of kinematic quantum states are $|{\bf x}, \phi\rangle:= |{\bf x}\rangle|\phi \rangle$, where  \\$|{\bf x}\rangle:=|\mu_1 j_1, \mu_2 j_2, \mu_3 j_3\rangle$
and $|\phi \rangle$ are the eigenvectors of the scalar field, such that
\be\label{kinba}
\langle {\bf x'}, \phi '|{\bf x}, \phi \rangle= \delta_{\bf x',\bf x} \delta(\phi ', \phi ).
\ee
We can therefore write (\ref{ga}) in this basis as
\be\label{green1}
\langle {\bf x}, \phi |\Psi_{phys}) = \sum_{\bf x'}\int d\phi '\; A({\bf x}, \phi ; {\bf x'}, \phi ')\Psi_{kin}({\bf x'},
\phi '),
\ee
where the Kernel $A({\bf x}, \phi ; {\bf x'}, \phi ')$ is given by
\be\label{green2}
A({\bf x}, \phi ; {\bf x'}, \phi ')=\int d\alpha \langle {\bf x}, \phi |e^{i\alpha \hat C}|{\bf x'},\phi '\rangle.
\ee
\section{The Path Integral Approach}
We shall follow here the path integral approach,  based on \cite{klein} and developed for a timeless framework in \cite{ash1}, which consists essentially in replacing the transition function in
Feynman's formalism by the Kernel $A({\bf x}_f, \phi _f; {\bf x}_I, \phi _I)$, where the subscripts $f$ and $I$ denote the final and initial
states of the system, and regarding the constraint operator $\exp(i\alpha \hat C)$ in (\ref{green2}) in a purely mathematical sense as a Hamiltonian with evolution time equal to one. That is, $e^{i\alpha \hat C}= e^{it\hat H}$
where $\hat H=\alpha \hat C$ and $t=1$.
Emulating now the standard Feynman construction, we decompose the fictitious evolution into $N$ infinitesimal evolutions of length $\lambda=\frac{1}{N+1}$. Thus we get
\begin{align}\label{pathint}
\langle {\bf x}_f, \phi _f|e^{i\alpha \hat C}|{\bf x}_I,\phi _I\rangle= \sum_{{\bf x}_{N},\dots,{\bf x}_{1}}\int d\phi _{N}\dots d\phi _1\times \langle {\bf x}_{N+1}, \phi _{N+1}|e^{i\lambda\alpha \hat C}|{\bf x}_{N},\phi _{N}\rangle\dots \langle {\bf x}_{1}, \phi _{1}|e^{i\lambda\alpha \hat C}|{\bf x}_{0},\phi _{0}\rangle,
\end{align}
where
$\langle {\bf x}_f, \phi _f|\equiv \langle {\bf x}_{N+1}, \phi _{N+1}|$ and $|{\bf x}_I,\phi _I\rangle\equiv |{\bf x}_{0},\phi _{0}\rangle$.
If we now consider in detail the particular n-th term in (\ref{pathint}) we can readily derive expressions for the remaining other terms. Thus, with $\hat C$ as given by (\ref{qconst1})
 we get
\begin{align}\label{pathint2}
\langle {\bf x}_{n+1}, \phi _{n+1}|e^{i\lambda\alpha \hat C}|{\bf x}_{n},\phi _{n}\rangle &= \langle \phi _{n+1}|e^{-i\lambda\alpha \hat p^2_{\phi }}|\phi _n\rangle
\langle {\bf x}_{n+1}|e^{i\lambda\alpha \hat C_{grav}}|{\bf x}_{n}\rangle\nonumber\\
&=\left( \frac{1}{2\pi}\int dp_{n} e^{i\lambda\alpha p^2_{n}} e^{ip_{n}(\phi _{n+1} -\phi _{n}) }\right) \langle {\bf x}_{n+1}|e^{i\lambda\alpha \hat C_{grav}}|{\bf x}_{n}\rangle.
\end{align}

To evaluate the gravitational constraint factor above note that, to order one in  $\lambda =\frac{1}{N+1}$ and for $N\gg 1$ we have
\be\label{pathint3}
\langle {\bf x}_{n+1}|e^{i\lambda\alpha \hat C_{grav}}|{\bf x}_{n}\rangle\approx \delta_{{\bf x}_{n+1}, {\bf x}_{n}} +
i\lambda\alpha \langle {\bf x}_{n+1}|\hat C_{grav}|{\bf x}_{n}\rangle + {\mathcal O}(\lambda^2).
\ee
Making use of (\ref{op7}), (\ref{op8}), as well as of (\ref{Lambda2}) -(\ref{vmult}) we see that there are 16 terms conforming the transition function $\langle {\bf x}_{n+1}|\hat C_{grav}|{\bf x}_{n}\rangle$. These
terms involve products of the unitaries and/or their conjugates. Let us consider in detail the term
of the form
\be\label{pathint4b}
\langle {\bf x}_{(n+1)}|\hat V_{i} \hat V_{j} \hat U_{i} \hat U_{j} | {\bf x}_{(n)}\rangle = e^{-i\pi\varepsilon_{i} \varepsilon_{j} (\hat e_{i} \times\hat e_{j})\cdot \boldsymbol\theta}\; e^{-2\pi i (\varepsilon_{i} x_{(n)i} +\varepsilon_{j} x_{(n)j})}\langle {\bf x}_{(n+1)}-\mu_{i} \hat e_{i} -\mu_{j} \hat e_{j} |{\bf x}_{(n)}+\frac{1}{2}(\varepsilon_{i} \hat e_{i} +\varepsilon_{j} \hat e_{j})\times\boldsymbol\theta\rangle.
\ee
Now, as pointed out in Sec.2 we have associated the action of the translation group on itself as leading to an affine space with a discrete topology and with a coset decomposition
${\bf T}_3 =\sum^{\infty}_{j_1 ,j_2 , j_3=-\infty} (\mu_i j_i) {\hat e}_{i}$, where $j_{(l)i}\in {\mathbb Z}$ and
the ${\hat e}_{i}$ are the basic translations in ${\mathbb R}^{3}$.  The vectors $ {\bf x}_{(l)} = \sum_{i=1}^3(\mu_i j_{(l)i}){\hat e}_{i}\in{\bf T}_3 $
are elements of ${\mathbb R}^{3}$ as a group and the set $\Gamma:\{\mu_i j_{(l)i} \}$ form a 3-dimensional cell. This in turn led us ({\it cf} eqn. (\ref{intval})) to introduce a Kronecker inner product
for the space of these vectors.
Moreover, when using the GNS construction to derive the kinematic Hilbert space we were also led to require that the translations induced by the Unitary operators $\hat U_i$ and $\hat V_l$ should be related
in order that the ``reticulations" induced by any of them should coincide. We suggested there that such a coincidence could be achieved by establishing the relations (\ref{rel3}) and (\ref{rel4}). This can now be
verified directly by noting first that the arguments in the ``bra" vectors in (\ref{pathint4b}) are clearly integer multiples of the $\mu_i$ and so are the arguments of the ``ket" vectors provided the
following relations are satisfied:
\be\label{cond}
\frac{ \hat e_l \cdot [ (\varepsilon_{i} \hat e_{i} \pm \varepsilon_{j} \hat e_{j})\times\boldsymbol\theta]}{2 \mu_l}\in {\mathbb Z}.
\ee
These requirements are indeed identically satisfied by the relations (\ref{rel3}) and (\ref{rel4}) for all the entries in the transition function in (\ref{pathint3}).\\
Consequently
\begin{align}\label{pathint4}
\langle {\bf x}_{(n+1)}|\hat V_{i} \hat V_{j} \hat U_{i} \hat U_{j} | {\bf x}_{(n)}\rangle &= e^{-i\pi\varepsilon_{i} \varepsilon_{j} (\hat e_{i} \times\hat e_{j})\cdot \boldsymbol\theta}\; e^{-2\pi i (\varepsilon_{i} x_{(n)i} +\varepsilon_{j} x_{(n)j})}\prod_{l=1}^{3}\int^{\frac{1}{2\mu_l}}_{-\frac{1}{2\mu_l}} \mu_l dk_{(n)l}\nonumber\\
&\times e^{-2\pi i \mu_l k_{(n)l} (j_{(n+1)l}-j_{(n)l}) } e^{2\pi i k_{(n)l} [\mu_j \delta_{lj}+\mu_i \delta_{li}+\frac{1}{2}\hat e_l \cdot  (\varepsilon_{i} \hat e_{i} + \varepsilon_{j} \hat e_{j})\times\boldsymbol\theta]},
\end{align}
and making use of (\ref{repa}), (\ref{op6}) and (\ref{pathint4}) we
find that
\begin{align}\label{mast}
\sum_{i \neq j}\langle {\bf x}_{(n+1)}&|\hat p^i \hat p^{j} \hat a_{i} \hat a_{j} | {\bf x}_{(n)}\rangle =\frac{1}{2}\sum_{i<j} \frac{\cos[\pi\varepsilon_{i} \varepsilon_{j} (\hat e_{i} \times\hat e_{j})\cdot \boldsymbol\theta]} {\mu_i \mu_j \varepsilon_{i} \varepsilon_{j}}\int \mu_1 dk_{(n)1} \mu_2 dk_{(n)2} \mu_3 dk_{(n)3} \nonumber\\
& \times e^{-2\pi i \sum_{l=1}^3\mu_l k_{(n)l} (j_{(n+1)l}-j_{(n)l}) } \; \sin\left[ 2\pi\varepsilon_i\left( x_{(n)i}+ \frac{1}{2}\sum_{l}^3 k_{(n)l} \theta_{li}\right)  \right ]\\
&\sin\left[ 2\pi\varepsilon_j\left( x_{(n)j}+ \frac{1}{2}\sum_{l}^3 k_{(n)l} \theta_{lj}\right)  \right ]\sin(2\pi k_{(n)i} \mu_i)\sin(2\pi k_{(n)j} \mu_j).\nonumber
\end{align}

We can now use (\ref{mast}) as a master equation to derive the two terms of the gravitational constraint in (\ref{qconst1}). The resulting expression is
\begin{align}\label{mast2}
\langle {\bf x}_{n+1}|\hat C_{grav}|{\bf x}_{n}\rangle &=\prod_{l=1}^3 \int \mu_l d k_{(n)l} e^{-2\pi i  k_{(n)l} (\ x_{(n+1)l}- x_{(n)l})}\nonumber\\
&\times\Bigg\{\frac{1}{4} \sum_{i=1}^3 \frac{1}{\varepsilon_i^2 \mu_i^2}\sin^2\left[ 2\pi\varepsilon_i\left( x_{(n)i}+ \frac{1}{2}\sum_{l}^3 k_{(n)l} \theta_{li}\right)  \right ]
\sin^2(2\pi k_{(n)i}\mu_i)\nonumber \\
&- \frac{1}{2}\sum_{i<j}\cos[2\pi\varepsilon_i \varepsilon_j \boldsymbol\theta \cdot (\hat e_i \times \hat e_j )]\frac{1}{\varepsilon_i}\sin\left[ 2\pi\varepsilon_i\left( x_{(n)i}+ \frac{1}{2}\sum_{l}^3 k_{(n)l} \theta_{li}\right)  \right ]\nonumber\\
&\times \frac{1}{\varepsilon_j}\sin\left[ 2\pi\varepsilon_j\left( x_{(n)j}+ \frac{1}{2}\sum_{l}^3 k_{(n)l} \theta_{lj}\right)\right]  \frac{1}{\mu_i}\sin(2\pi k_{(n)i} \mu_i)\frac{1}{\mu_j}\sin(2\pi k_{(n)j} \mu_j )\Bigg\}
\end{align}


Inserting now (\ref{mast2}) into (\ref{pathint3}) and exponentiating, we have
\be\label{pathint5}
\langle {\bf x}_{(n+1)}|e^{i\lambda\alpha \hat C_{grav}}|{\bf x}_{(n)}\rangle= \prod_{l=1}^3 \int^{1/2\mu_l}_{-1/2\mu_l} \mu_l dk_{(n+1)l}\;
e^{-2\pi ik_{(n+1) l}({x}_{(n+1)l}-{ x}_{(n)l})} e^{i\lambda\alpha C_g ({\bf k}_{(n+1)},{\bf x}_{(n+1)}, {\bf x}_{(n)})} ,
\ee
where $C_g ({\bf k}_{(n+1)},{\bf x}_{(n+1)}, {\bf x}_{(n)})$ is the infinitesimal spectral contribution of the gravitational part of the constraint, given by the terms inside the braces in (\ref{mast2}).

Hence, substituting each of the corresponding infinitesimal amplitude terms in (\ref{pathint5}) into the gravitational part of (\ref{pathint}) yields
\be\label{panthint6}
\langle {\bf x}_f |e^{i\alpha \hat C_g}|{\bf x}_I\rangle = \prod_{l=1}^3 \;\left[\sum_{ j_{Nl}.. j_{1l}=-\infty}^\infty\right]\prod_{n=0}^N \int^{\frac{1}{2\mu_l }}_{-\frac{1}{2\mu_l}}
\mu_l dk_{(n+1)l}\; e^{-2\pi ik_{(n+1) l}\mu_l ({j}_{(n+1)l}-{ j}_{(n)l})} e^{i\lambda\alpha C_g ({\bf k}_{(n+1)},{\bf x}_{(n+1)}, {\bf x}_{(n)})}.
\ee
Now, in order to arrive at an expression involving a proper continuous path integral, we follow the procedure described in \cite{klein} and consider first the amplitude (\ref{panthint6}) for the case of
no constraint. We then have
\be\label{panthint7}
\langle {\bf x}_f |{\bf x}_I\rangle_{0} := \prod_{l=1}^3 \;\left[\sum_{ j_{Nl}.. j_{1l}=-\infty}^\infty\right]\left[\prod_{n=0}^N \int^{\frac{1}{2 }}_{-\frac{1}{2}}
 d\bar k_{(n+1)l}\right]\; e^{-2\pi i\sum_{n=0}^N\bar k_{(n+1) l}({j}_{(n+1)l}-{ j}_{(n)l})},
\ee
where we have absorbed the $\mu_l$'s in the integrations by redefining $\bar k_{(n+1)l}:=\mu_l k_{(n+1)l}$. \\
Note next that the summation in the exponential in (\ref{panthint7}) can be reordered as follows:
\be
\sum_{n=0}^N\sum_{l=1}^3 \bar k_{(n+1) l}({j}_{(n+1)l}-{ j}_{(n)l}) =\sum_{l=1}^3 \left[\bar k_{(N+1)l}j_{(f)l} -  \bar k_{(1)l}j_{(I)l} -\sum_{n=1}^N j_{(n)l}(\bar k_{(n+1)l}-\bar k_{(n)l})\right].
\ee
Substituting this expression back into (\ref{panthint7}) and using the Poisson formula, we arrive at
\begin{align}\label{panthint8}
\langle {\bf x}_f |{\bf x}_I\rangle_{0} := \prod_{l=1}^3 \left[\prod_{n=0}^N \int^{\frac{1}{2 }}_{-\frac{1}{2}}
 d\bar k_{(n+1)l}\right]e^{-2\pi i(\bar k_{(N+1)l}j_{(f)l}-\bar k_{(1)l}j_{(I)l})}&\prod_{n=1}^N \left[\sum_{m_{(n)l} =-\infty}^\infty \delta(\bar k_{(n+1)l}- \bar k_{(n)l}+m_{(n)l})\right],\nonumber\\
 &\quad\quad\quad\quad\quad\quad\quad\quad\quad\quad m_{(n)l} \in \mathbb Z.
\end{align}
Using now the Fourier integral representation of the Dirac delta function we alternatively can write
\begin{align}\label{panthint9}
\langle {\bf x}_f |{\bf x}_I\rangle_{0} :&= \prod_{l=1}^3 \left[\prod_{n=0}^N \int^{\frac{1}{2 }}_{-\frac{1}{2}}
 d\bar k_{(n+1)l}\right]e^{-2\pi i(\bar k_{(N+1)l}j_{(f)l}-\bar k_{(1)l}j_{(I)l})}\nonumber\\
&\times\prod_{n=1}^N \left[\int_{-\infty}^\infty d\bar q_{(n)l} \right]\sum_{m_{(n)l} =-\infty}^{\infty}\left( e^{-2\pi i\sum_{n=1}^N \bar q_{(n)l} (\bar k_{(n+1)l}- \bar k_{(n)l}+m_{(n)l})}\right),
\end{align}
where the unitless $\bar q_{(n)l}\in \mathbb R $.
Noting that the integers $-\infty\leq m_{(n)l}\leq \infty$ in the sum in the above exponential can be absorbed into the variables $\bar k_{(n)l}$ for $1\leq n \leq N$ so their range of integration is
extended to $(-\infty, \infty) $, we therefore can write

\begin{align}\label{panthint10}
\langle {\bf x}_f |{\bf x}_I\rangle_{0}=& \prod_{l=1}^3  \int^{\frac{1}{2 }}_{-\frac{1}{2}} d\bar k_{(N+1)l} e^{-2\pi i \bar k_{(N+1)l}j_{(f)l}} \prod_{n=1}^N
\left[\int_{-\infty}^\infty d\bar k_{(n)l} \right]e^{2\pi i \bar k_{(1)l}j_{(I)l}} \nonumber\\
&\times\prod_{n=1}^N \left[\int_{-\infty}^\infty d\bar q_{(n)l} \right]e^{2\pi i\sum_{n=1}^N \bar q_{(n)l} (\bar k_{(n+1)l}- \bar k_{(n)l})}.
\end{align}
Rearranging once more the summation in the exponential above, we obtain
\be\label{panthint11}
\langle {\bf x}_f |{\bf x}_I\rangle_{0}= \prod_{l=1}^3  \int^{\frac{1}{2 }}_{-\frac{1}{2}} d\bar k_{(N+1)l}  \prod_{n=1}^N
\left[\int_{-\infty}^\infty d\bar k_{(n)l} \right] \int_{-\infty}^\infty d\bar q_{(n)l} e^{-2\pi i \sum_{n=0}^{N} \bar k_{(n+1)l}(\bar q_{(n+1)l}- \bar q_{(n)l}) }.
\ee
after denoting the end-points as $\bar q_{(N+1)l}:=j_{(f)l}$ and $\bar q_{(0)l}:= j_{(I)l)}$.

Comparing now the amplitude (\ref{panthint11}) with (\ref{panthint6}), we note that the sum over the discrete variables $j_{(n)l}\in\mathbb Z$  in (\ref{panthint6}) is replaced by the continuous
$\bar q_{(n)l} \in\mathbb R$ in (\ref{panthint11}). Therefore we can introduce in the summation of the exponential in (\ref{panthint11}) the symbol (the term inside the braces
of (\ref{mast2}))of the constraint operator $\hat C_g$ acting on the spectral representation of the infinitesimals $\langle {\bf x}_{n+1}||{\bf x}_{n}\rangle_0$, after replacing the $j_{(n)l}$ discrete variables by the
$q_{(n)l}$ continuous ones. Thus
\begin{align}\label{panthint12}
\langle {\bf x}_f| e^{i\alpha \hat C_g}|{\bf x}_I\rangle
= &\prod_{l=1}^3 \int^{\frac{1}{2 }}_{-\frac{1}{2}} d\bar k_{(N+1)l}\prod_{n=1}^N
\left[\int_{-\infty}^\infty d\bar k_{(n)l} \int_{-\infty}^\infty d\bar q_{(n)l}\right]   \nonumber\\
&\times e^{-i \sum_{n=0}^{N}\left[ 2\pi \bar k_{(n)l}(\bar q_{(n+1)l}- \bar q_{(n)l}) -\alpha(\frac{1}{N+1}) C_g (\bar k_{(n)l}, \bar q_{(n)l},\mu,\varepsilon)\right]}.
\end{align}
Making next use of the above expression in the evaluation of (\ref{pathint}) and (\ref{pathint2}) yields
\begin{align}\label{pathint13}
\langle {\bf x}_f, \phi _f|e^{i\alpha \hat C}|{\bf x}_I,\phi _I\rangle =&\prod_{l=1}^3 \int^{\frac{1}{2 }}_{-\frac{1}{2}} d\bar k_{(N+1)l}\prod_{n=1}^N
\left[\int_{-\infty}^\infty d\bar k_{(n)l} \int_{-\infty}^\infty d\bar q_{(n)l}\right] e^{-2\pi i \bar k_{(1)l}(\bar q_{(1)l}-j_{(I)l})}\nonumber\\
&\times \frac{1}{(2\pi)^N}e^{-2\pi i \bar k_{(N+1)l}(j_{(f)l}-\bar q_{(N)l})}\left[\prod_{n=1}^{N} \int d\phi _{(n)}\right]\left[\prod_{n=0}^{N} \int dp_{\phi(n)}\right] e^{-iS_N},
\end{align}
with
\be\label{action6}
S_N = -\lambda\sum_{n=0}^{N} \left[p_{\phi(n)} \left(\frac{\phi _{(n+1)} -\phi _{(n)}}{\lambda}\right) -2\pi \sum_{l=1}^3 \bar k_{(n)l} \left(\frac{\bar q_{(n+1)l} - \bar q_{(n)l}}{\lambda}\right)
+ \alpha\Bigg (\frac{1}{2}p^2_{\phi(n)}+C_g(\bar k_{(n+1)l}, \bar q_{(n)l},\mu,\varepsilon) \Bigg)\right].
\ee

The last step in the path integral procedure consists in letting $\lambda = \Delta \tau$ so that (\ref{action6}) reads
\be\label{action7}
S_N = \sum_{n=0}^{N} \Delta\tau\Bigg [-p_{\phi(n)} \left(\frac{\phi _{(n+1)} -\phi _{(n)}}{\Delta\tau}\right) +2\pi \sum_{l=1}^3 \bar k_{(n)l} \left(\frac{\bar q_{(n+1)l} - \bar q_{(n)l}}{\Delta\tau}\right)
-\alpha\Bigg (\frac{1}{2}p^2_{\phi(n)}+ C_g(\bar k_{(n+1)l}, \bar q_{(n)l},\mu,\varepsilon) \Bigg)\Bigg].
\ee
Further taking the limit $N\to \infty$
\be\label{action7b}
S:= \lim_{N\to \infty} S_N = \int_{\tau=0}^{\tau=1} d\tau\left[- p_{\phi} \dot\phi  + 2\pi {\bf\bar k}(\phi )\cdot {\bf{\dot{\bar q}}}(\phi ) -\alpha(\frac{1}{2}p_\phi^2 + C_g({\bf\bar k}(\phi ),{\bf\bar q}(\phi ),\mu,\varepsilon))  \right]
\ee
and varying $p_\phi$ results in the equation of motion $\dot\phi = -\alpha p_\phi$. Write now
\be\label{hamalt}
d\tau =d\phi\left(\frac{d\tau}{d\phi}\right)=\frac{d\phi}{\dot\phi},
\ee
so that
\begin{align}\label{action7c}
S =& \int_{\phi (\tau=0)}^{\phi (\tau=1)} d\phi \left[ 2\pi {\bf\bar k}(\phi )\cdot {\bf{\dot{\bar q}}}(\phi ) -p_{\phi} -\left(\frac{\alpha}{\dot\phi}\right)\left(\frac{1}{2}p_{\phi}^2 + C_g({\bf\bar k}(\phi ),{\bf\bar q}(\phi ),\mu,\varepsilon)\right)  \right]\nonumber\\
&=\int_{\phi (\tau=0)}^{\phi (\tau=1)} d\phi \left( 2\pi {\bf\bar k}(\phi )\cdot {\bf{\dot{\bar q}}}(\phi ) -\left[p_{\phi} -\left(\frac{1}{p_\phi}\right)\left(\frac{1}{2}p_{\phi}^2 + C_g({\bf\bar k}(\phi ),{\bf\bar q}(\phi ),\mu,\varepsilon)\right)\right]  \right),
\end{align}
where from here on ``dot" means differentiation with respect to the internal time $\phi $. With this reparametrization the term in the square brackets in the second equality above is the Hamiltonian of the system,
so (\ref{action7c}) can be written as

\be\label{action7d}
S = \int_{\phi (\tau=0)}^{\phi (\tau=1)} d\phi \left[ 2\pi {\bf\bar k}(\phi )\cdot {\bf{\dot{\bar q}}}(\phi ) - H \right],
\ee
where
\be\label{hamalt2}
H =\frac{p_\phi}{2} -\left(\frac{1}{p_\phi}\right) C_g({\bf\bar k}(\phi ),{\bf\bar q}(\phi ),\mu,\varepsilon)=E,
\ee
and the energy $E$ is a constant of motion.
By combining the above different contributions to the action the explicit form of this Hamiltonian is given by
\begin{align}\label{totalS}
H= &\left(\frac{1}{p_{\phi}}\right)
\Bigg[\frac{p^2_{\phi}}{2} + \frac{1}{4}\sum_{i=1}^3 \frac{1}{\varepsilon_i^2 \mu_i^2}\sin^2
\left[2\pi\varepsilon_i \mu_i \left( \bar q_i (\phi ) -\frac{1}{2}\sum_{l=1}^3 \frac{\theta_{il}\bar k_l}{\mu_i \mu_l}   \right)\right]\sin^2 (2\pi \bar k_i)\nonumber\\
& -\frac{1}{2}\Bigg\{\sum_{\substack{i,j=1\\i< j}}^3 \cos[2\pi \varepsilon_i \varepsilon_j \boldsymbol\theta \cdot (\hat e_i \times \hat e_j)]\frac{1}{\varepsilon_i \mu_i}
\sin \left[2\pi\varepsilon_i \mu_i \left( \bar q_i (\phi ) -\frac{1}{2}\sum_{l=1}^3 \frac{\theta_{il}\bar k_l}{\mu_i \mu_l}   \right)\right]\sin (2\pi \bar k_i)\\
&\times \frac{1}{\varepsilon_j \mu_j}
\sin \left[2\pi\varepsilon_j \mu_j \left( \bar q_j (\phi ) -\frac{1}{2}\sum_{l=1}^3 \frac{\theta_{jl}\bar k_l}{\mu_j \mu_l}   \right)\right]\sin (2\pi \bar k_j) \Bigg\}\Bigg]\nonumber.
\end{align}

In order to get a further physical insight on the terms in (\ref{totalS}), consider the expectation value of the operator $\hat a_i$ as defined in (\ref{repa}):

\begin{align}
\langle\Psi| \hat a_i |\Psi\rangle=-&  \frac{1}{2i\varepsilon_i}\langle \Psi| U_i - U^{\dag}_i |\Psi\rangle=-\frac{1}{2i\varepsilon_i}\sum_{j_1, j_2 , j_3}\langle
\Psi| U_i - U^{\dag}_i|{\bf x}\rangle\langle {\bf x}|\Psi\rangle =\nonumber\\
&=-\frac{1}{2i\varepsilon_i}\sum_{j_1, j_2 , j_3}
\Bigg[ e^{-2\pi i\varepsilon_i x_i} \Psi^{\ast}\left({\bf x} + \frac{1}{2}\varepsilon_i \hat e_i \times \boldsymbol\theta\right)
 -e^{2\pi i\varepsilon_i x_i} \Psi^{\ast}\left({\bf x} - \frac{1}{2}\varepsilon_i \hat e_i \times \boldsymbol\theta\right)\Bigg]\Psi({\bf x}) \label{expec}
\end{align}

Recalling now ({\it cf} (\ref{FT2})) that

\be\label{FT2b}
\Psi({\bf x}) = \prod_{l=1}^3 \int_{-\frac{1}{2\mu_l}}^{\frac{1}{2\mu_l}}dk_l \;\Phi(k_l) e^{-2\pi i k_l \mu_lj_l},
\ee
and substituting into (\ref{expec}), we get
\begin{align}\label{expec2}
\langle \Psi|\hat a_i |\Psi\rangle = &\frac{1}{\varepsilon_i}\sum_{j_1,j_2, j_3}\int d^3 k^{\prime} \int d^3 k \; \Phi^{\ast} ({\bf k}) \Phi ({\bf k}^{\prime}) e^{-2\pi i\sum_l \mu_l j_l(k_l - k^{\prime}_l)  }\nonumber\\
&\times\sin\left[ 2\pi \varepsilon_i \mu_i \left(j_i +\frac{{\bf k}\cdot(\hat e_i \times\boldsymbol\theta)}{2\mu_i}\right)\right].
\end{align}
Consider now the scalar
\be\label{scalar}
\langle \Psi|\Psi \rangle= \sum_{j_1, j_2, j_3}\langle \Psi|{\bf x}\rangle\langle {\bf x}|\Psi\rangle, \;\;\; {\bf x} = \sum_{l=1}^3 \mu_l j_l \hat e_l,
\ee
which, making again use of (\ref{FT2b}) and the Poisson sum formula results in the spectral decomposition
\be\label{scalar2}
\langle \Psi|\Psi \rangle=\int d^3 k^{\prime} \int d^3 k \; \Phi^{\ast} ({\bf k}^{\prime}) \Phi ({\bf k})\int_{-\infty}^{\infty} d^3 \bar q \;e^{-2\pi i\sum_l \mu_l \bar q_l(k_l - k^{\prime}_l)}.
\ee
Comparing (\ref{expec2}) with (\ref{scalar2}) we see that we can identify the function
\be\label{symba}
(a_i)_{symb}:=\frac{1}{\varepsilon_i}\sin\left[ 2\pi \varepsilon_i \mu_i \left(\bar q_i +\frac{{\bf k}\cdot(\hat e_i \times\boldsymbol\theta)}{2\mu_i}\right)\right]
\ee
as the symbol of $\hat a_i$ acting on the spectral representation of $\langle\Psi|\Psi\rangle$, with $j_l = x_l /\mu_l$ going to the continuum limit $j_l \to \bar q_l$. Hence we can infer
from (\ref{totalS}) that this same function is the symbol of $\hat a_i (\phi )$.
In particular, note that since noncommutativity is dominant at distances of the order of a Planck length where the sine function can be well approximated by its argument,  it is natural to identify
the dimensionless quantities $\bar k_i$ and
\be
\bar Q_i =\left(\bar q_i +\frac{1}{2\mu_i} {\bf k}\cdot (\hat e_i \times\boldsymbol\theta) \right)=
\left( \bar q_i (\phi ) -\frac{1}{2}\sum_{l=1}^3 \frac{\theta_{il}\bar k_l}{\mu_i \mu_l}\right),
\ee
which satisfy the twisted Poisson bracket algebra $\{\bar Q_i ,\bar Q_j\}=(2\pi)^{-1}\frac{\theta_{ij}}{\mu_i \mu_j}$ and $ \{\bar Q_i, \bar k_j \} = \frac{1}{2\pi} \delta_{ij} $,
in the effective Hamiltonian of the path integral formulation. Moreover, recalling that $Q_i =\mu_i\bar Q_i$ and $\bar k_j =\mu_j k_j $ we have that the above  expressions when appropriately dimensioned as dynamical coordinates of the trajectories  and their respective canonical conjugate momenta, become
\be\label{comm8}
\{Q_i , Q_j\}=(2\pi)^{-1}\theta_{ij}\quad \text {and} \quad \{ Q_i,  k_j \} = \frac{1}{2\pi} \delta_{ij} ,
\ee
 which coincide with their Poisson brackets given by a Moyal $\star$-product algebra.

Making next use of these variables and defining
\be\label{altham}
\chi_i := \frac{1}{\varepsilon_i \mu_i} \sin(2\pi \varepsilon_i \mu_i \bar Q_i ) \sin(2\pi \bar k_i),
\ee
and
\begin{align}\label{altham3}
\alpha &:=\cos[2¹\pi \varepsilon_1 \varepsilon_2 \boldsymbol\theta \cdot (\hat e_1 \times \hat e_2)] \nonumber\\
\beta&:=\cos[2¹\pi \varepsilon_1 \varepsilon_3 \boldsymbol\theta \cdot (\hat e_1 \times \hat e_3)]\\
\gamma&:= \cos[2¹\pi \varepsilon_2 \varepsilon_3 \boldsymbol\theta \cdot (\hat e_2 \times \hat e_3)],\nonumber
\end{align}
we can rewrite (\ref{totalS}) as
\be\label{altham4}
 H =\left(\frac{1}{p_\phi}\right) \left[\frac{1}{2}p_{\phi}^2 +
\frac{1}{4}\left[\chi_1\left(\chi_1 -\alpha\chi_2 -\beta \chi_3 \right) +\chi_2\left(\chi_2 -\alpha\chi_1 -\gamma \chi_3 \right)+\chi_3\left(\chi_3 -\beta\chi_1 -\gamma \chi_2 \right)\right]\right]=E.
\ee
Furthermore, if we now implement the Hamiltonian constraint strongly, that is to say \\$\left(\frac{1}{2}p_{\phi}^2 + C_g({\bf\bar k}(\phi ),{\bf\bar q}(\phi ),\mu,\varepsilon)\right)=0$, we have from
(\ref{hamalt2}) that
 $E=p_\phi$. Hence
\be\label{altham5}
\frac{p^2_\phi}{2}-C_g({\bf\bar k}(\phi),{\bf\bar q}(\phi ),\mu,\varepsilon)=E p_\phi= p^2_\phi
\ee
and
\be\label{altham6}
\left[\frac{1}{2}p_{\phi}^2 +
\frac{1}{4}\left[\chi_1\left(\chi_1 -\alpha\chi_2 -\beta \chi_3 \right) +\chi_2\left(\chi_2 -\alpha\chi_1 -\gamma \chi_3 \right)+\chi_3\left(\chi_3 -\beta\chi_1 -\gamma \chi_2 \right)\right]\right]=0.
\ee

\section{Asymptotics for the Noncommutative Dynamics}

The dynamics of our system is given in the stationary phase approximation by the solution of the equations:
\begin{align}\label{eqm1}
{\dot{\bar k}_i} =& -\frac{1}{2 p_\phi}\cos\left(2\pi\varepsilon_i \mu_i \bar Q_i\right)\sin\left(2\pi\bar k_i\right)R_i, \;\;\;i=1,2,3
\end{align}
where
\be\label{eqm1b}
R_1 := \left(\chi_1 -\alpha \chi_2 -\beta\chi_3 \right),\;\;\; R_2 := \left(\chi_2 -\alpha \chi_1 -\gamma\chi_3 \right),\;\;\; R_3 := \left(\chi_3 -\beta\chi_1 -\gamma\chi_2 \right),
\ee
\be\label{eqm2}
{\dot{\bar Q}_i}=\left(\frac{1}{p_\phi}\right)\left(\frac{1}{2\varepsilon_i \mu_i }\sin\left(2\pi\varepsilon_i \mu_i \bar Q_i \right)\cos\left( 2\pi \bar k_i \right) R_i
- \sum_{j\neq i}^3\frac{ \theta_{ij}}{\mu_i \mu_j} {\dot{\bar k}_j}\right).
\ee
Now, to be able to assert the dynamical behavior of the observables ${\bar Q}_i$ and ${\bar k}_i$, let us first make use of (\ref{altham}) to derive  explicitly the time
derivative of ${\bar k}_i$. We get
\be\label{qm3}
{\dot{\bar k}_i} = \left(\frac{1}{2\pi}\right)\frac{d}{d\phi }\left(\frac{\varepsilon_i \mu_i\chi_i}{\sin\left(2\pi\varepsilon_i \mu_i \bar Q_i \right)}\right)
{\Bigg [ 1-\left(\frac{\varepsilon_i \mu_i\chi_i}{\sin\left(2\pi\varepsilon_i \mu_i \bar Q_i \right)}\right)^2 \Bigg ]^{-1/2}},\;\;\; i=1,2,3.
\ee
Substituting (\ref{eqm1}) into the left hand side of (\ref{qm3}) results in
\be\label{qm4}
\left( \frac{\pi}{p_\phi}\right)  \cos(2\pi\varepsilon_i \mu_i \bar Q_i) R_i= \frac{d}{d\phi }\cosh^{-1} \left(\frac{\sin\left(2\pi\varepsilon_i \mu_i \bar Q_i \right)}{\varepsilon_i \mu_i\chi_i} \right),  \;\;\; i=1,2,3
\ee
and by integrating yields
\be\label{qm5}
\sin\left(2\pi\varepsilon_i \mu_i \bar Q_i \right)= \varepsilon_i \mu_i \chi_i \cosh \left [\frac{ \pi}{ p_\phi} \int^{\phi (\tau)}_{\phi (I)} d\phi  \cos\left(2\pi\varepsilon_i \mu_i \bar Q_i \right)R_i \: + B_i \right],  \;\;\; i=1,2,3
\ee
where $\phi (I)$ is the inner-time at the boundary conditions, the constant of integration $B_i$ is the evaluation
\be
B_i= \cosh^{-1}\left(\frac{\sin\left(2\pi\varepsilon_i \mu_i \bar Q_i \right)}{\varepsilon_i \mu_i\chi_i} \right)|_{\phi(I)},
\ee
and the sign of the left hand side of (\ref{qm5}) has to be taken consistent with the sign of the $\chi_i$ on the right hand side. As we show in the paragraph following equation (\ref{qm15}) the $\chi_i$
can be taken consistently to be positive for all times, thus it follows from (\ref{qm5}) that the symbol of $\hat a_i$ acting on the spectral representation of $\langle\Psi|\Psi\rangle$ has to satisfy the inequality
\be\label{qm6}
\frac{|\sin\left(2\pi\varepsilon_i \mu_i \bar Q_i \right)|}{\varepsilon_i}\geq \mu_i \chi_i,
\ee
as it is also evident from (\ref{altham}).

Next, in order to derive the time evolution of the $\bar k_i$'s we make use of  (\ref{eqm1}) to write
\be\label{qm7}
\frac{\dot{\bar k}_i}{\sin\left( 2\pi \bar k_i \right) } =-\left(\frac{1}{2 p_\phi}\right)\cos\left(2\pi \varepsilon_i \mu_i \bar Q_i \right)R_i
\ee
which integrates (for i=1,2,3) to

\begin{align}\label{qm9}
\tan(\pi\bar k_{i} (\phi(\tau))  =& \tan(\pi \bar k_{i}(\phi(B))) \left(\exp \Bigg[-\frac{\pi}{p_\phi} \int^{\phi (\tau)}_{\phi (I)} d\phi  \cos\left(2\pi\varepsilon_i \mu_i \bar Q_i \right) R_i \Bigg ]\right).
\end{align}
To complete this stage of our analysis we need to consider the dynamical evolution of the $\chi_i$'s into which the Hamiltonian constraint is decomposed. Note, by the way, that these quantities turn out to be
constants of the motion in the limit of zero noncommutative symbol. Let us then multiply both sides of (\ref{eqm2}) by $\cot (2\pi \varepsilon_i \mu_i \bar Q_i )$. We get
\begin{align}\label{qm10}
2\pi \varepsilon_i \mu_i \cot(2\pi \varepsilon_i \mu_i \bar Q_i ){\dot{\bar Q}_i} =& \frac{\pi}{p_\phi} \cos (2\pi \varepsilon_i \mu_i \bar Q_i ) \cos(2\pi\bar k_i) R_i -\nonumber\\
&-\left(\frac{2\pi}{p_\phi}\right)\varepsilon_i \cot(2\pi \varepsilon_i \mu_i \bar Q_i )\sum_{j\neq i}^3 \frac{\theta_{ij}}{\mu_j}{\dot{\bar k}_j},
\end{align}
which can be re-expressed as
\be\label{qm11}
\frac{d}{d\phi } \ln \left(\sin (2\pi \varepsilon_i \mu_i \bar Q_i )\right)= -2\pi \cot(2\pi\bar k_i){\dot{\bar k}_i}-(2\pi) \sum^3_{j\neq i}\theta_{ij} \frac{\varepsilon_i}{ \mu_j} \cot(2\pi
\varepsilon_i \mu_i \bar Q_i ){\dot{\bar k}_j},
\ee
or, passing the first term on the right above as a differential to the left and making use of (\ref{altham}) and (\ref{eqm1}), as
\be\label{qm12}
\frac{d}{d\phi } \ln \left( \varepsilon_i \mu_i \chi_i \right) =\pi\sum_{j\neq i}\varepsilon_i \varepsilon_j \theta_{ij} \chi_j R_j
\cot(2\pi \varepsilon_i \mu_i \bar Q_i )\cot(2\pi \varepsilon_j \mu_j \bar Q_j ).
\ee
Multiplying both sides of (\ref{qm12}) by $\chi_i R_i$ for $i=1,2,3$ we can eliminate the terms on the right by adding the resulting three equations. Thus we get
\be
 R_1 \dot{\chi}_1 +  R_2 \dot{\chi}_2 + R_3 \dot{\chi}_3=0.
\ee


As a check of consistency note that this result equally follows from differentiating (\ref{altham6}) with respect to the inner time, since it is easy to show that
\be\label{qm15}
\frac{d}{d\phi} \left(p_\phi^2 = -\frac{1}{2} (\chi_1 R_1 + \chi_2 R_2 +\chi_3 R_3)\right)\Longrightarrow R_1 \dot{\chi}_1 +  R_2 \dot{\chi}_2 + R_3 \dot{\chi}_3=0.
\ee

The above makes only sense provided the signs of the $\chi_i$'s in (\ref{altham6}) and therefore inside the parenthesis in (\ref{qm15}) are such that the equation makes sense. To establish this
we note that since $p_{\phi}$ is a constant of the motion and evidently can not be chosen as zero, we are then required that $\frac{1}{2} (\chi_1 R_1 + \chi_2 R_2 +\chi_3 R_3)$ be negative definite at any time $\phi $.
It is easy to verify that this implies that none of the  $\chi_i$'s can be zero at any time. Indeed, assume that $\chi_1=0$, then $p_{\phi}^2 = -\frac{1}{2}\left[ (\chi_2 -\gamma\chi_3)^2 + \chi^2_3 (1-\gamma^2)\right]$,
 which is clearly impossible
unless $\chi_2$ and $\chi_3$ are imaginary which is evidently not so as seen from (\ref{altham}). An entirely similar argument applies if we were to set $\chi_2$ or $\chi_3$ equal to zero since in this cases
we would get as inconsistencies $p_{\phi}^2 = -\frac{1}{2}\left[ (\chi_1 -\beta\chi_3)^2 + \chi^2_3 (1-\beta^2)\right]$ and
 $p_{\phi}^2 = -\frac{1}{2}\left[ (\chi_1 -\alpha\chi_2)^2 +(\chi_2)^2 (1-\alpha^2 )\right]$ which is again impossible for $\chi_i$'s real. Hence all three
$\chi_i$'s must be either positive or negative definite.

It is not difficult to show that the $\chi_i $'s can be chosen to be positive at a particular time. For instance by requiring that the $R_i$ be negative at that time. That they can indeed be chosen positive
for all times can be seen when integrating (\ref{qm12}).
The resulting integral equations are exponentials of the form
\begin{align}\label{qm18}
 \chi_i( \phi (\tau))=& \chi_i( {\phi (B)})\times
\exp\bigg [\pi\sum^3_{j\neq i}\varepsilon_i \varepsilon_j \theta_{ij} \int_{\phi(I)}^{\phi(\tau)} \chi_j R_j
\cot(2\pi \varepsilon_i \mu_i \bar Q_i )\cot(2\pi \varepsilon_j \mu_j \bar Q_j ) d\phi\bigg],
\end{align}
which are therefore always positive and can never reach zero according to our previous considerations.\\

Next, based on the developments in Sec.V leading to equation (\ref{symba}) for the symbols of the operators $\hat a_i$, we can define the volume of the Bianchi I Universe as the product of these symbols, {\it i.e.} as:

\begin{align}\label{expval2}
{\mathcal V}_{symb}= \prod_{i=1}^{3}(a_i)_{symb}=\frac{1}{\varepsilon_1 \varepsilon_2 \varepsilon_3}\Bigg[\sin(2\pi\varepsilon_1 \mu_1\bar Q_1)\sin(2\pi\varepsilon_2\mu_2\bar Q_2)\sin(2\pi\varepsilon_3\mu_3\bar Q_3)\Bigg].
\end{align}
That this definition is reasonable follows from the fact that the $\hat a_i$ are noncommutative and can not be used as simultaneous observables and also because in the limit of commutativity we have that
\be\label{symba2}
\lim_{\varepsilon\to 0}({\mathcal V}_{symb})=\prod_{i=1}^3 (2\pi\mu_i \bar Q_i) .
\ee
Moreover, so far the quantities $\varepsilon_i$, $\mu_i$ were introduced in the $C^\ast$-algebra discussed in Section II in order to account primarily for the proper dimensions in equations (\ref{Lambda})-(\ref{vmult}) describing its
realization, we can go one step further in our analysis by interpreting $\varepsilon_i$ and $\mu_i$ as scale parameters describing the different stages of evolution of the dynamical system. We shall now express them as scale factors by writing
\be\label{scf1}
\varepsilon_i = \frac{\bar\varepsilon_i}{L_i},
\ee
where $\bar\varepsilon_i$ is a constant and $L_i$ is in units of length and magnitude depending on the corresponding scale at which the evolving universe is considered. Correspondingly, since at a scale where noncommutativity is expected to be dominant the $\varepsilon_i$ and the
$\mu_i$ are related by equations (\ref{rel3}) and (\ref{rel4}), we will have that
\be\label{scf1b}
n_j \varepsilon_i \mu_i = n_i \varepsilon_j \mu_j, \quad\quad i \neq j
\ee
and
\be\label{scf2}
\mu_1 =\frac{n_1}{2} \frac{\bar\varepsilon_2}{ L_2}\lambda^2_P \bar\theta_3 , \;\;\;\; \mu_2 =\frac{n_2}{2} \frac{\bar\varepsilon_1}{
L_1}\lambda^2_P \bar\theta_3 , \;\;\;\;  \mu_2 =\frac{n_3}{2} \frac{\bar\varepsilon_1}{L_1}\lambda^2_P \bar\theta_2 ,
\ee
(and consistent with our previous notation bared quantities are dimensionless throughout).
Thus, in particular, we find that
\be\label{scf3}
\varepsilon_1 \mu_1 = \frac{n_1}{2} \frac{\bar\varepsilon_1\bar\varepsilon_2}{L_1 L_2}\lambda^2_P \bar\theta_3 .
\ee
Noting now that at the Planck length scale the area in the plane perpendicular to the vector $\hat e_3$ is related to the symbol of the commutator $[\hat a_1, \hat a_2]$ we see that when substituting
(\ref{scf3}) into (\ref{uncert6}) that
\be\label{scf42a}
(s_3)_{0} \approx 2\pi  \boldsymbol\theta\cdot(\hat e_1 \times\hat e_2),
\ee
and similarly for the two other planes we have
\be\label{scf42b}
(s_2)_{0} \approx 2\pi  \boldsymbol\theta\cdot(\hat e_3 \times\hat e_1),\quad\quad (s_1)_{0} \approx 2\pi  \boldsymbol\theta\cdot(\hat e_2 \times\hat e_3),
\ee
so that the magnitude of the minimal area of the Bianchi I universe is determined by the noncommutativity and is proportional to the square of the Planck length in magnitude value, similar to expressions obtained by other approaches in different contexts.

One more indicator on the actual values to be assigned to the scale factors $L_i$ in (\ref{scf1}) can be derived from the conceptually expected noncommutativity of the algebras describing physical processes occurring at
distances of the order of the Planck length. In mathematical terms this would be equivalent to express the range of validity of the noncommutativity in our equations by introducing a smooth cutoff function in the $\varepsilon_i$
of (\ref{scf1}) with compact support when the universe conforms a region of radial dimensions of the order of Planck lengths. To this
end we make use of Theorem 1.4.1 in \cite{hor}, which shows that a test function
$\psi_i \in C_0^\infty (X)$ of compact support, in an open set in $\Bbb R ^3$, can be found with $0\leq \psi_i\leq 1$ so that $\psi_i=1$ in a neighborhood of a compact subset $K$ of $X$. The regularization $\psi_i$ of
$\varepsilon_i$ is thus obtained by the convolution
\be\label{scf5}
\psi_i:= \chi_{K_{2\rho}} \ast \varphi_{\rho} \in C_0^\infty ( K_{3\rho}),
\ee
where $\chi_{K_{2\rho}}$ is the characteristic function of
\be\label{scf6}
K_{2\rho}:= \{ y, |x-y| \leq 2\rho, \text {for some}\; x \in K\},
\ee
and $\varphi_{\rho}$ is the mollifier
\be\label{scf7}
\varphi_{\rho} (y)=\rho^{-3} \exp{\Bigg[-\frac{1}{(1-\frac{|y|^2}{\rho^2})}\Bigg]}.
\ee
It therefore follows from (\ref{scf5}) and (\ref{scf6}) that for radii of the order of $10\lambda_P$ noncommutativity will be supported in a ball of radius $30\lambda_P$, so we can identify $\bar\varepsilon_i$ with
$\psi_i$, which is equal to one inside the ball and zero outside, and  use $L_i \approx 30\lambda_P$ for the effective
regularization cutoff of the noncommutativity terms in our evolution equations; {\it i.e.}

\be\label{scf7b}
\bar\varepsilon_i=\psi_i =\int_{B_{\bar L_i}}dy \;\delta(y-y_0) =\begin{cases} 1&\text{for $\;y_0< \frac{L_i}{\lambda_P} =30$}\\
0 &\text {for $ y_0 \geq 30$}
\end{cases}
\ee

Thus for $\bar Q_i$ such that $(a_i)_{symb} < 30$ the argument in the left hand side of (\ref{qm5}) becomes,
after making use of (\ref{scf3}) and (\ref{scf7b}), $2\pi\varepsilon_i \mu_i \bar Q_i \approx \frac{\pi n_i \bar\varepsilon_i \bar\varepsilon_j \bar\theta_k \bar Q_i}{900}= \frac{n_i \pi\bar\theta_k \bar Q_i}{900}$ (where i,j,k are cyclically ordered),
while for $\bar Q_i$ such that $(a_i)_{symb}\geq 30$, since $\bar\varepsilon_i =0$, we then have
\be\label{scf8}
\lim_{\bar\varepsilon_i\to 0}\left(\frac{\sin (2\pi\varepsilon_i \mu_i \bar Q_i)}{\varepsilon_i \mu_i}\right)= 2\pi \bar Q_i.
\ee
 Consequently  above this cutoff scale we need to replace (\ref{qm5}), (\ref{qm9})and (\ref{altham}) by

\be\label{scf9}
\bar Q_i (\phi (\tau))= \frac{ \chi_i(\phi (L_i))}{2\pi} \cosh \left [\frac{ \pi}{ p_\phi}R_i \Big(\phi (\tau) - \phi  (L_i)\Big) +B_i (L_i) \right],  \;\;\; i=1,2,3
\ee
where here $B_i (L_i)$ is the evaluation
\be\label{scf9b}
B_i (L_i)= \cosh^{-1}\left(\frac{\sin\left(2\pi\varepsilon_i \mu_i \bar Q_i \right)}{\varepsilon_i \mu_i\chi_i} \right)|_{\phi(L_i)},
\ee
\be\label{scf10}
\tan(\pi\bar k_{i} (\phi(\tau))  = \tan(\pi \bar k_{i}(\phi(L_i))) \left(\exp \Bigg[-\frac{\pi}{p_\phi} R_i \Big(\phi (\tau) -\phi( L_i) \Big )  \Bigg ]\right),
\ee
\be\label{scf11}
\chi_i(\phi(\tau)) = 2\pi \bar Q_i (\phi(\tau)) \sin\Big(2\pi\bar k_i(\phi(\tau))\Big),
\ee
in our evolution calculations, with $R_i$ and $\chi_i$ becoming constants of motion due to the effective absence of noncommutativity beyond this cutoff.

Now observe that (\ref{symba2}) already states the role of the quantities $2\pi\mu_i\bar Q_i$ as the physical configuration variables in the limit $\varepsilon\rightarrow0$, which in turn imply that  volume and areas in the commutative regime are measured in multiples of an elementary volume $(2\pi)^3\mu_1\mu_2\mu_3$ and elementary areas $(2\pi)^2\mu_i\mu_j$ respectively. Because this can only be the reminiscence of the minimal areas (\ref{scf42a}) and (\ref{scf42b}) from the noncommutative regime then
\be\label{scf11a}
(2\pi)^2\mu_1\mu_2=2\pi\theta_3,\quad(2\pi)^2\mu_2\mu_3=2\pi\theta_1,\quad(2\pi)^2\mu_1\mu_3=2\pi\theta_2,
\ee
or equivalently
\be\label{scf11b}
\frac{\theta_3}{\mu_1\mu_2}=\frac{\theta_1}{\mu_2\mu_3}=\frac{\theta_2}{\mu_1\mu_3}=2\pi.
\ee

By making use of (\ref{scf11b}) along with (\ref{rel3}) and (\ref{rel4}) it is straightforward to show that $n_1=n_2=n_3$ and equation (\ref{scf1b}) reduces to
\be\label{scf11c}
\varepsilon_1 \mu_1 = \varepsilon_2 \mu_2=\varepsilon_3 \mu_3.
\ee

In order to implement these notions so that the system can be faithfully evolved with the noncommutative equations inside the noncommutative region and with the commutative ones beyond the cutoff, we will require compatible solutions for both scenarios. This compatibility can be achieved through the selection of appropriate boundary values occurring at the cutoff region, which may be obtained by analyzing the behavior of $\dot\chi_i$.

Because one of the main differences between the noncommutative system and the commutative one is the constancy of all the $\chi_i$'s or equivalently $\dot\chi_i=0$ in the commutative case, this also establishes a criteria to determine when and how the noncommutative system can follow the commutative evolution beyond the cutoff. By using eq. (\ref{qm12}) it is immediate that
\be\label{scf12}
\dot\chi_i =\pi\sum_{j\neq i}\frac{\theta_{ij}}{\mu_i\mu_j} R_j
\cos(2\pi \varepsilon_i \mu_i \bar Q_i )\cos(2\pi \varepsilon_j \mu_j \bar Q_j )\sin(2\pi\bar k_i)\sin(2\pi\bar k_j).
\ee

From the previous expression we can obtain the values $\bar Q_i,\bar k_i$ for which $\dot\chi_i=0$, which are clearly given by
\be\label{scf13}
\bar Q_i=(-1)^r\frac{2r+1}{4\varepsilon_i\mu_i},\quad\bar k_i=\frac{s}{2},\quad r,s\in\mathbb{Z},\quad i=1,2,3
\ee
where the factor $(-1)^r$ guarantees the positivity of the symbol associated to $\hat a_i$.

However, because it is precisely when valued at (\ref{scf13}) that $\dot{\bar k}_i=0$ and the symbols of $\hat a_i$ reach their maximum and their rate of change becomes zero, there is ambiguity in continuing the evolution of the system beyond such values with expressions (\ref{scf9}) and (\ref{scf10}). To circumvent this difficulty we have to look for more adequate boundary values where the system can be said to be expanding or contracting, but where we still have $\dot\chi_i\approx0$ at any chosen order.

By looking at intervals centered in (\ref{scf13}) we may define the set of boundary conditions
\be\label{scf14}
\bar Q_i(0)=(-1)^r\frac{2r+1}{4\varepsilon_i\mu_i}+\frac{\zeta_i}{2\pi},\quad\bar k_i(0)=\frac{s}{2}+\frac{\delta_i}{2\pi},\quad0<|\zeta_i|\leq\frac{\pi}{2\varepsilon_i\mu_i},\quad0<|\delta_i|\leq\frac{\pi}{2},
\ee
where expanding solutions correspond to ${\zeta_i}<0$ and contracting ones to ${\zeta_i}>0$. After substituting this in (\ref{scf12}) we get
\be\label{scf15}
\dot\chi_i(0) =\pi\sum_{j\neq i} \frac{\theta_{ij}}{\mu_i\mu_j} R_j
\sin(\varepsilon_i \mu_i \zeta_i )\sin(\varepsilon_j \mu_j \zeta_j  )\sin(\delta_i)\sin(\delta_j).
\ee

Noting from (\ref{altham}) that $|\chi_i|\leq\frac{1}{\varepsilon_i\mu_i}$ and consequently $|R_i|\leq\frac{3}{\varepsilon_i\mu_i}$ and using $|\sin(\alpha)|\leq|\alpha|$, we can establish an upper bound for the absolute value of $\dot\chi_i(0)$ and using (\ref{scf11b}) yields
\be\label{scf16}
|\dot\chi_i(0)| =\bigg|2\pi^2\sum_{j\neq i}R_j
\sin(\varepsilon_i \mu_i \zeta_i )\sin(\varepsilon_j \mu_j \zeta_j  )\sin(\delta_i)\sin(\delta_j)\bigg|\leq6\pi^2\varepsilon_i\mu_i\sum_{j\neq i}
|\zeta_i| |\zeta_j||\delta_i||\delta_j|,
\ee

For an upper bound $M\in\mathbb{R}^+$ such that
\be\label{scf18}
6\pi^2\varepsilon_i\mu_i\sum_{j\neq i}|\zeta_i| |\zeta_j||\delta_i||\delta_j|\leq M,
\ee
the inequalities can be solved to obtain
\be\label{scf19}
|\zeta_i||\delta_i|\leq\sqrt{\frac{M}{12\pi^2\varepsilon_i\mu_i}},
\ee
which can be further relaxed if all the $\chi_i$'s are chosen to have the same sign and so $|R_i|\leq\frac{2}{\varepsilon_i\mu_i}$, in which case
\be\label{scf20}
|\zeta_i||\delta_i|\leq\sqrt{\frac{M}{8\pi^2\varepsilon_i\mu_i}}.
\ee

Finally we need to enforce the cutoff condition in the interval of validity of $\zeta_i$. This is done directly from demanding
\be\label{scf21}
\frac{1}{\varepsilon_i}\sin(2\pi\varepsilon_i\mu_i\bar Q_i(0))\geq L_i,
\ee
or equivalently
\be\label{scf22}
\frac{1}{\varepsilon_i}\cos(\varepsilon_i\mu_i|\zeta_i|)\geq L_i,
\ee
which for our case where $\varepsilon_i\mu_i|\zeta_i|\leq\frac{\pi}{2}$ also implies
\be\label{scf23}
|\zeta_i|\leq \frac{1}{\varepsilon_i\mu_i}\arccos (\varepsilon_iL_i).
\ee

Together, the inequalities (\ref{scf22}) and (\ref{scf23}) provide the refinement for the admissible intervals of values for $\zeta_i$ and $\delta_i$ expressed now as
\be\label{scf24}
0<|\zeta_i|\leq\frac{1}{\varepsilon_i\mu_i}\arccos(\varepsilon_iL_i),\quad0<|\delta_i|\leq\sqrt{\frac{M}{8\pi^2\varepsilon_i\mu_i}}\frac{1}{|\zeta_i|}.
\ee

This criteria provides with the full description of the system below and above the cutoff where from expression (\ref{scf8}) the matching boundary conditions at the cutoff region must satisfy
\be
\begin{split}
(a_i)_{symb}(0)&=\frac{1}{\varepsilon_i}\sin(2\pi\mu_i\varepsilon_i\bar Q_i(0))=2\pi\mu_i\bar Q_i(0),\\
\chi_i(0)&=\frac{1}{\varepsilon_i\mu_i}\sin(2\pi\mu_i\varepsilon_i\bar Q_i(0))\sin(2\pi\bar k_i(0))=2\pi\bar Q_i(0)\sin(2\pi\bar k_i(0)),
\end{split}
\ee
which implements the change of physical variables when going from below the cutoff to the region above.

In this sense any trajectory governed by the noncommutative algebra evolution of expressions (\ref{eqm1}) and (\ref{eqm2}), with boundary values (\ref{scf14}) and (\ref{scf24}) at the cutoff region, obeys a compatible commutative evolution (to order $M$) outside the Planckian region determined by (\ref{scf9}-\ref{scf11}).\\

\begin{figure}[h!]
  \centering
  \includegraphics[width=12cm]{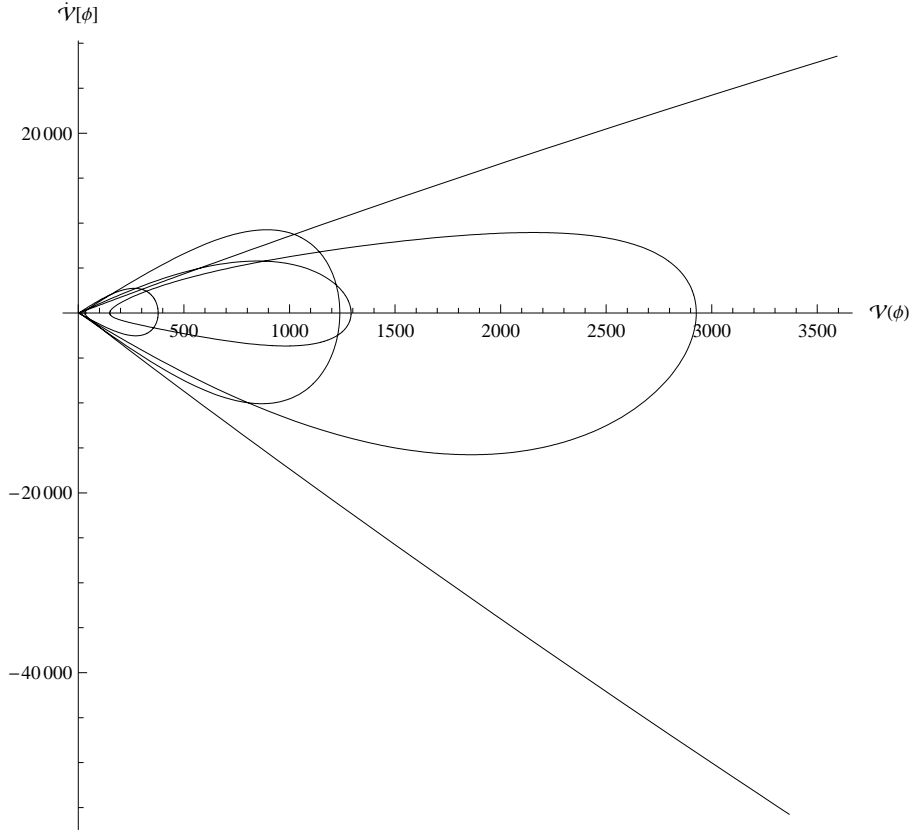}\\
  \caption{Phase-space plot of the volume with visible transition from an open collapsing orbit (lower branch) to periodic orbits connecting various invariant tori ending with an open expanding orbit (upper branch).}\label{10}
\end{figure}

The results just obtained  can be further explained as follows. The system has a 6-dimensional phase-space, of which a suitable parametrization of a projection is the 2-dimensional plot
$({\mathcal V}_{symb} , \dot{\mathcal V}_{symb})$ shown in Fig.(\ref{10}) (this phase-space diagram applies to the case discussed in section 8 with reference to Fig.(\ref{5}) ) . This figure shows a monotone orbit followed by an oscillatory behavior emerging into a new expanding orbit. Even though the quantities
$\varepsilon_i , \mu_j$   are linked by the fundamental physics $\theta_{ij}$, strictly from a differential equations point of view we can consider $\theta_{ij}= 0$ with $\varepsilon_i , \mu_j \neq 0$. Then
when $\theta_{ij}= 0$, the $R_i$ are constant and the equations (which follow from multiplying (\ref{altham}) by $R_i$)
\be\label{scf25}
R_i \chi_i=\left(\frac{R_i}{\varepsilon_i \mu_i}\right) \sin(2\pi \varepsilon_i \mu_i \bar Q_i) \sin(2\pi\bar k_i) =\text {const.}
\ee
provide a family of invariants of the system. thus in this formulation the universe will oscillate in a quasi-periodic way. Now, when $\theta_{ij}\neq 0$
the tori are subjected to the corresponding Hamiltonian perturbation.

Consequently the unperturbed orbits have now periods which depend on the amplitude (this can be seen simply by quadrature using (\ref{scf25}) for each degree of freedom. Moreover, as the orbits approach the origin in the $\bar Q_i$ variables the period
becomes longer, since this is a hyperbolic point. Then the classical KAM results (\cite{moser}) guarantee the existence of nearby invariant tori for a large (in measure)
set of unperturbed tori. In the actual behavior of the solutions we have that, generically, the basic periodic solution of the $i^{th}$ degree of freedom pics up two
more periods due to the interaction with the two other phases. When the invariant tori come close to the separatrix the basic orbit has a long period. These corrections
will cause the oscillations. Furthermore, since the basic solutions have long periods, the resulting orbits become very sensitive (as the numerics in the following Section shows) to the parameters and initial conditions. When considering the implications of this behavior in the evolution of the volume, we would expect a relatively
fast contracting orbit away from the saddle point merging with a long period resulting thus in a periodic oscillation caused by the noncommutativity and merging
again (due to the integrability of the commutative problem) with the expanding solution.\\
It is important to recall that this behavior is not special but generic and is expected for any noncommutative model with an integrable structure in the commutative limit.
We therefore can conclude from the above that generically the noncommutative scenario and its induced evolution of the the invariants (\ref{scf25}),
 produces multiple solutions and effective noncommutative lattice structures as a consequence of the cosmology dynamics.

\section{Numerical Solutions}

In order to provide consistent values for the parameters in the equations and for appropriate initial conditions in the interesting parameter regimes described qualitatively in the previous section, let us now recall equations (\ref{rel3}) and (\ref{rel4}) which may be written as $\mu_i =\frac{n_i}{2}\varepsilon_j \theta_k $ with the indices $i,j,k$ ordered cyclically. Expressing the  above equation in units of Planck lengths we have
\be\label{ad1}
\bar\mu_i \lambda_P =\frac{n_i}{2}\frac{\bar\varepsilon_j \bar\theta_k}{\bar L_j} \lambda_P ,
\ee
where, as defined previously, bared symbols denote their magnitude and $\bar L_j$ is the magnitude of the scale factor of the $\varepsilon_j$.
Let us next consider the behavior of the two terms in the right of equation (\ref{eqm2}). In the Planck region the scale magnitude of $\bar L_j$ is of the order of a Planck length so also setting the scale magnitude $n_i$ of $\mu_i$ equal to a few Planck lengths we have that $\mu_i =
\varepsilon_j \theta_k \approx 1\lambda_P=\mathcal{O}(\lambda_p)$.
Consequently $\mu_{i} \varepsilon_{i}$ is of the order of one in this case. Applying a similar reasoning to the expression  $\frac{ \theta_{ij}}{\mu_i \mu_j}$
 we get that
\be\label{ad2} \mu_1 \mu_2 \approx \frac{4\lambda_P^2}{ \bar\varepsilon_1 \bar\varepsilon_2 \bar\theta_2 \bar\theta_3 }= \mathcal{O}(\lambda^2_P),
\ee
 which makes it consistent with (\ref{scf11b}) and, since for calculation  simplicity we are taking the tensor of noncommutativity to be of the same magnitude for all three planes, the second term on the right of equation (\ref{eqm2}) turns out to be commensurate with the first.

\begin{figure}[h!]
  \centering
  \includegraphics[width=12cm]{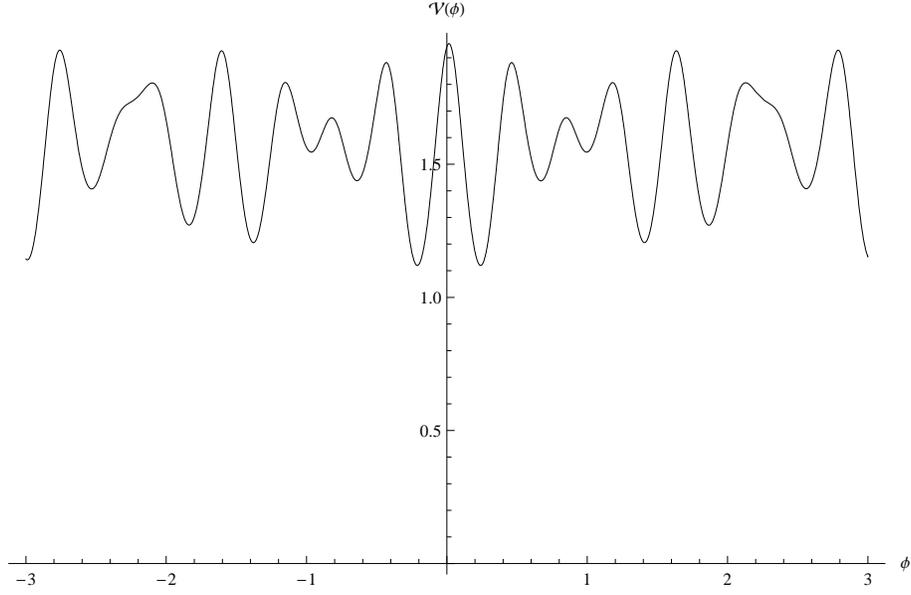}\\
  \caption{For $\varepsilon_i = 0.8(\lambda_P)^{-1}$, solutions for the Volume (with initial conditions for the radii symbols of order $\lambda_p$) display oscillatory behavior. Maxima and minima are always within the same order of magnitude and the system is confined to Planckian volume scales. }\label{1}
\end{figure}

\begin{figure}[h!]
  \centering
  \includegraphics[width=12cm]{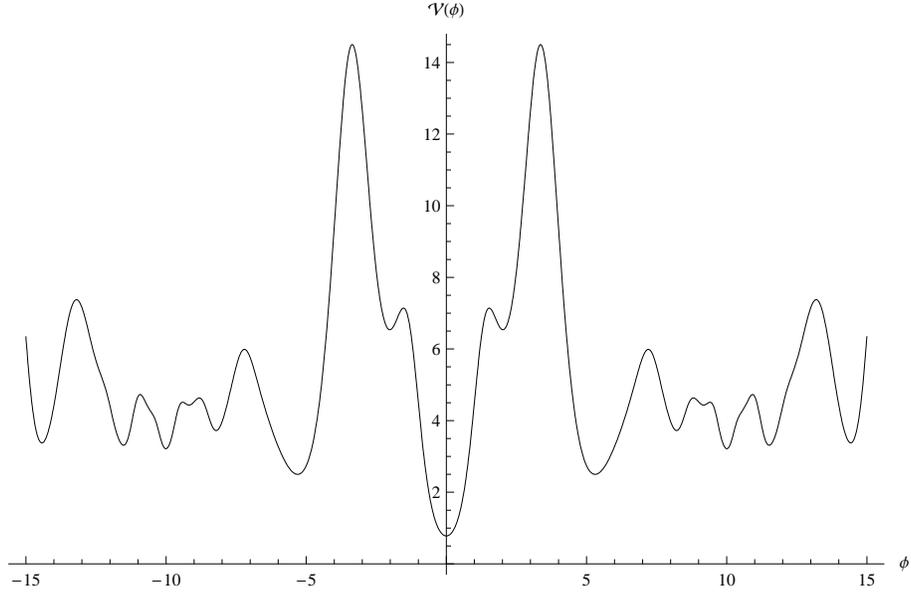}\\
  \caption{Solution for $\varepsilon = 0.4(\lambda_P)^{-1}$ . For smaller $\epsilon_i$ the system has access to bigger volumes and constructive interference among the independent symbols of the radii allows the formation of maxima of orders of magnitude greater than the minima. For values of $\epsilon_i<1/L_i$ these maxima eventually reach the cutoff region where the solutions are governed by the commutative regime and Eqs. (7.140)-(7.143). }\label{2}
\end{figure}

To illustrate the possible scenarios and how markedly they depart in the noncommutative case from classical (and non-classical) solutions, consider then the strongly noncommutative solutions of (\ref{expval2}) which occur when the noncommutative force term described above is commensurate with the first term in (\ref{eqm2}) at all times. As mentioned, this corresponds to values of $\varepsilon_i$ such that $\varepsilon_i\mu_i$ is of order one. Fig.\ref{1} and Fig.\ref{2} constitute examples of this regime, with evident similar properties, obtained for numerical values of $\varepsilon_i=0.8(\lambda_p)^{-1}$ and $\varepsilon_i=0.4(\lambda_p)^{-1}$ respectively. As neither of the solutions can reach the scales that would make noncommutative effects negligible the solutions are confined to Planckian scale volumes.

\begin{figure}[h!]
  \centering
  \includegraphics[width=12cm]{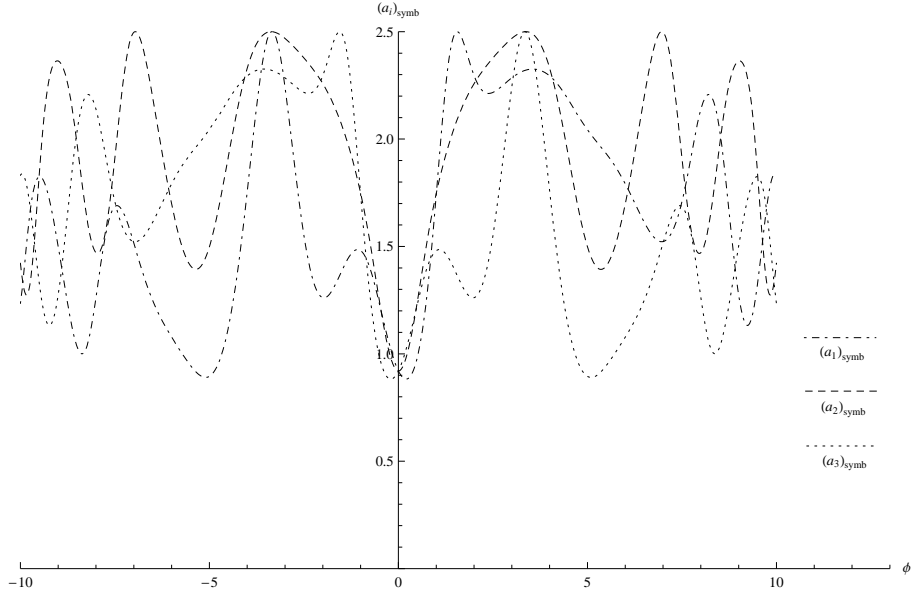}\\
  \caption{The independent symbols $(a_1)_{symb}, (a_2)_{symb}, (a_3)_{symb}$, associated to the volume in Fig.\ref{2}, display complex evolutions due to the noncommutative force term that mixes interactions in the three independent directions}\label{7}
\end{figure}

Although similar, the system in Fig.\ref{2} is seen to evolve more diversely than in Fig.\ref{1} with global minima and maxima now differing by orders of magnitude. The irregular oscillatory behavior is in both cases the product of the noncommutative force term acting as a drive, modulating the frequencies of the solutions of the independent symbols of the radii of the universe, as can be better observed in Fig.\ref{7} where the three independent symbols $(a_i)_{symb}$ associated to the volume in Fig.\ref{2} have been plotted. This shows explicitly that it is the noncommutativity the agent which eventually
drives the universe to scales past the Planckian scale through the smooth cutoff.

\begin{figure}[h!]
  \centering
  \includegraphics[width=12cm]{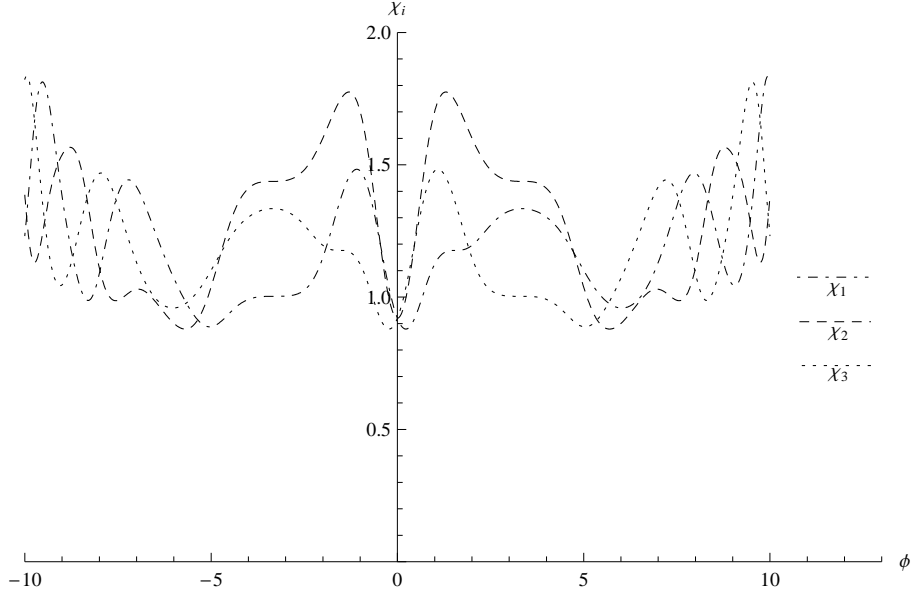}\\
  \caption{Plot of $\chi_1,\chi_2,\chi_3$ associated to the volume in Fig.\ref{2} where the
  approximately adiabatic regions around $\phi\approx\pm3.3$ correspond to the global maxima seen for the volume.}\label{8}
\end{figure}

\begin{figure}[h!]
  \centering
  \includegraphics[width=12cm]{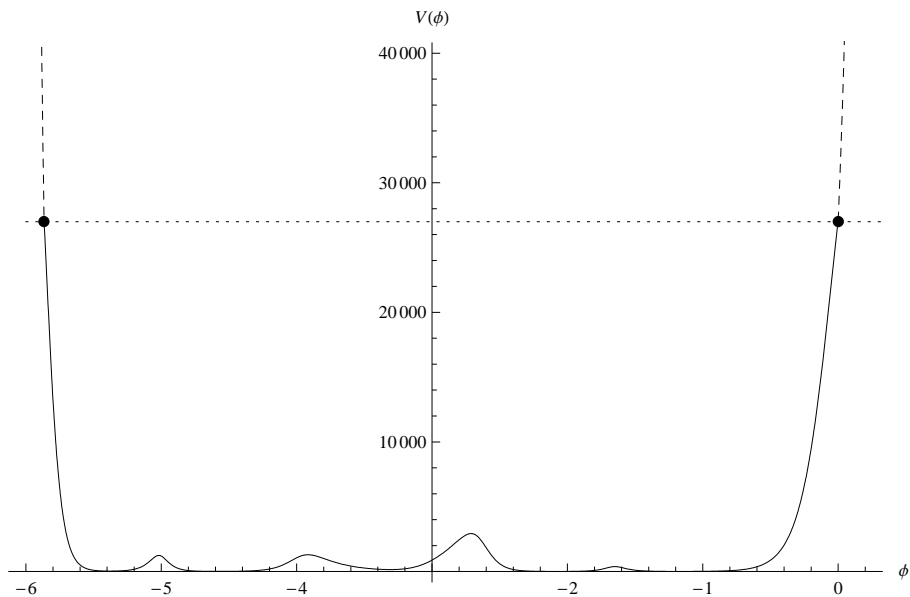}\\
  \caption{Collapsing and expanding solution for $\varepsilon_i = 0.031 (\lambda_P)^{-1}$ . The noncommutative evolution (solid), compatible with the boundary values of a collapsing solution (dashed) that enters from the left of the figure, remains inside the noncommutative region for a finite period of time before constructive interference brings the system back to the commutative region  expanding away from the cutoff.   }\label{5}
\end{figure}

By analyzing the $\chi_i$ variables, which in the commutative case are constants of motion and therefore can be interpreted as action variables, it is observed from Fig.\ref{8} that their behavior in the Planckian regime is not adiabatic and noncommutativity is not simply a perturbation. In fact, the abrupt changes of these variables are associated to minima of the volume where noncommutative effects are stronger, whereas approximately adiabatic regions correspond to maxima of the volume and such regions become more and more dominant at larger scales. It is then that the evolution of the system can continue along commutative states, which is the basis for our selection of boundary values at the cutoff, as confirmed by the following cases.

Thus, let us now consider the evolution when approaching the cutoff
from below, {\it i.e.} near $\bar L_i= 30$ then, by virtue of
(\ref{scf8}), the first term on the right of (\ref{eqm2}) becomes
$\pi \bar Q_i \cos(2\pi \bar k_i) R_1$ with $R_i$ given by
(\ref{eqm1b}) with $\alpha=\beta=\gamma=0$ and the $\chi_i$ becoming
constants of motion. On the other hand, after observing that
(\ref{ad2}) is independent of scales, and therefore the coefficients
of ${\dot{\bar k}_j}$ are again of order one and the second term
becomes negligible relative to the first one so the evolution beyond
this stage is given by equations (\ref{scf9})-(\ref{scf11});
In this case $\bar Q_i \approx \bar q_i$.\\
Moreover, observe that
 $\sum_{j\neq i}^3\frac{ \theta_{ij}}{\mu_i \mu_j} {\dot{\bar k}_j}$  acts
 as a force with unitless "mass" $\frac{ \theta_{ij}}{\mu_i \mu_j}$
and unitless acceleration $ {\dot{\bar k}_j}$ driving the canonical
variables $\bar Q_i$ in a direction perpendicular to  their
$i^{th}$-components. This is made even more transparent when noting
that by setting the tensor of noncommutativity equal to zero in
(\ref{eqm2}) the $R_i$ become constants of motion and the remaining
first term becomes strictly oscillatory.

\begin{figure}[h!]
  \centering
  \includegraphics[width=12cm]{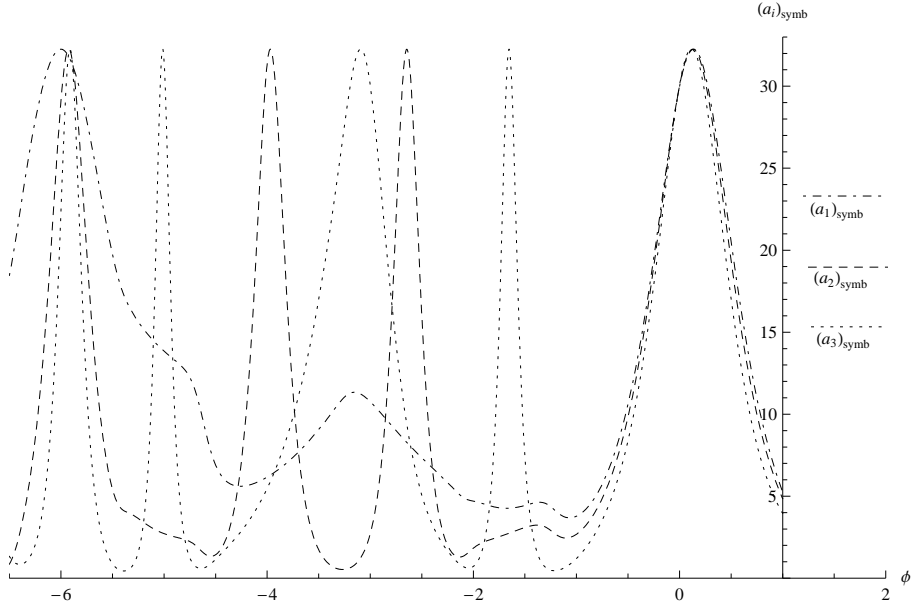}\\
  \caption{Independent symbols $(a_1)_{symb}, (a_2)_{symb}, (a_3)_{symb}$ for $\varepsilon= 0.031 (\lambda_P)^{-1}$. The constructive (resp. destructive) interference inside the noncommutative regime region leading to the evolution of the volume above (resp. below) the cutoff in (Fig.\ref{5}) is evidenced.   }\label{6}
\end{figure}

To exemplify this kind of solutions consider first the type of
bounce depicted in Fig.7. Here we have a scenario where a collapsing
trajectory (dashed) enters the noncommutative regime from the left,
leading to a noncommutative evolution (solid) below the cutoff,
where a number of noncommutative oscillations can be observed, until
the effects of the noncommutative force term bring the system to an
expansion phase such that it can reach the cutoff region and finally
continue along a continuous expansion. Fig.\ref{6} provides more
insight on the underlying interactions among the independent symbols
$(a_i)_{symb}$  that, due to the constructive and destructive
interferences, lead to the behavior of the volume shown inside the
noncommutative region.

\begin{figure}[h!]
  \centering
  \includegraphics[width=12cm]{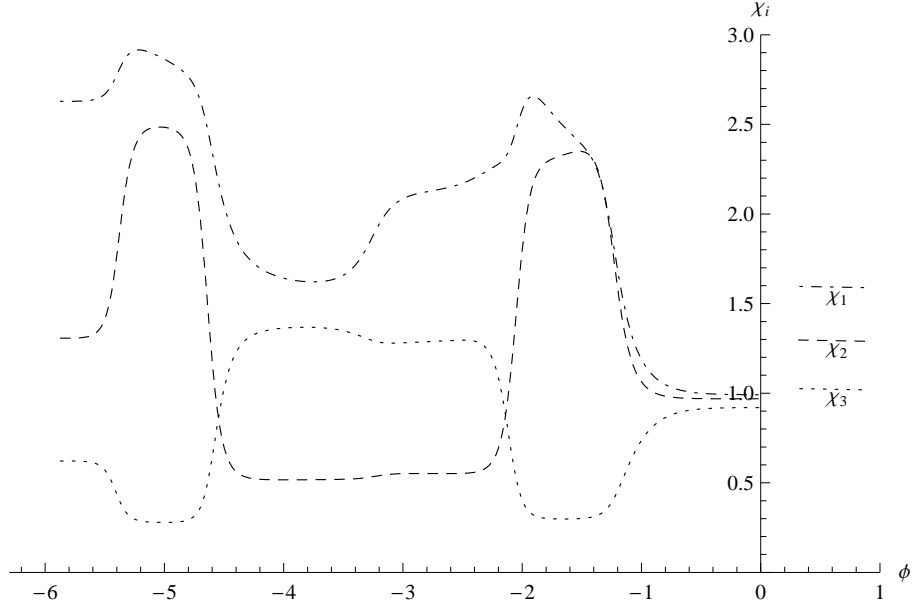}\\
  \caption{Plot of $\chi_1,\chi_2,\chi_3$ associated to the volume in Fig.\ref{5} where simultaneous regions of constant $\chi_i$ at the left and right of the figure lead to the asymptotic evolution of the volume beyond the cutoff.}\label{9}
\end{figure}

To finalize the discussion regarding this case compare the corresponding evolution of all
the $\chi_i$'s in Fig.\ref{9} with that of Fig.\ref{8} which confirms  the fact that at
larger scales the adiabatic regions become more dominant and, in particular, it is at
both extremes of Fig.\ref{9} that the system continues evolving for $\phi\gtrless0$ along those constant values of $\chi_i$.

\begin{figure}[h!]
  \centering
  \includegraphics[width=12cm]{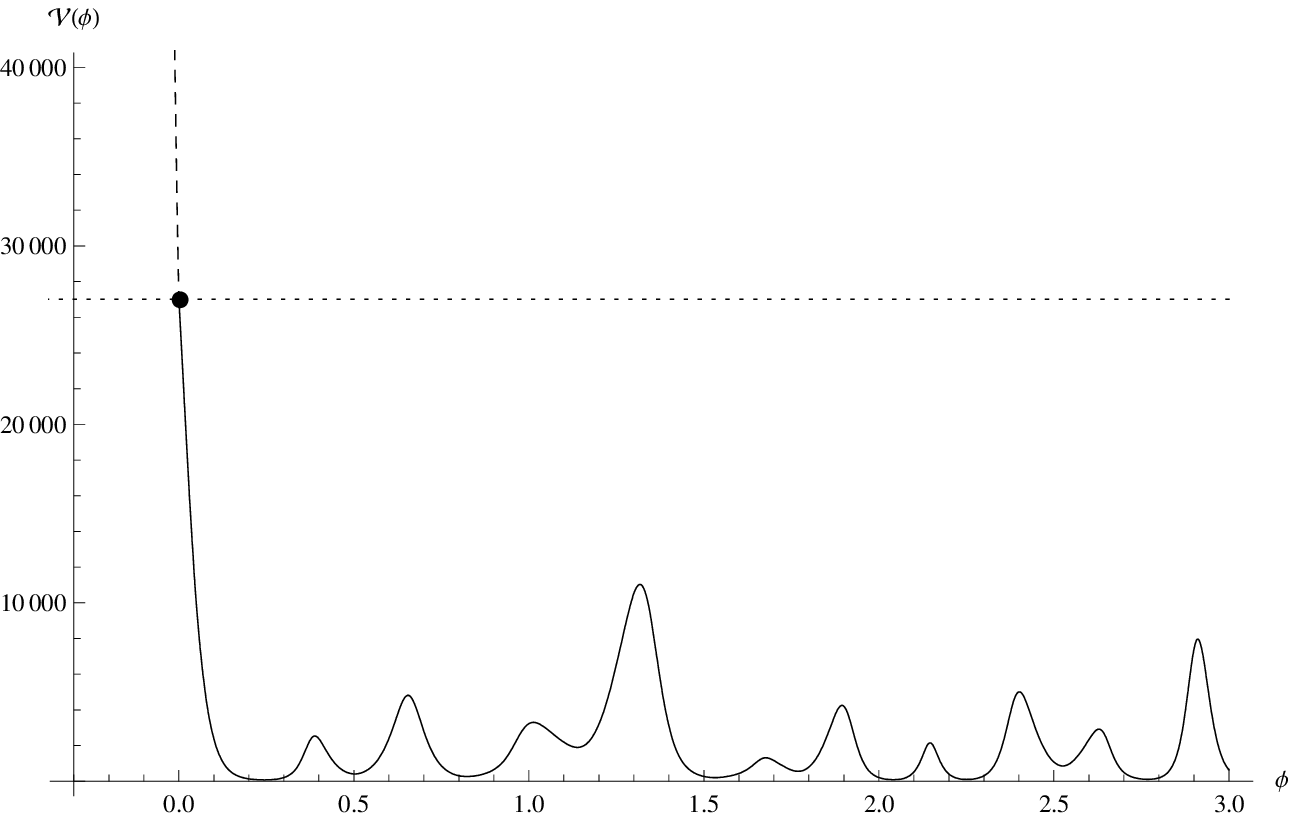}\\
  \caption{Collapsing solution for $\varepsilon_i= 0.031 (\lambda_P)^{-1}$ . The commutative regime solution (dashed) enters the noncommutative region through the cutoff (dotted) and continues below it along a noncommutative evolution with compatible boundary values (dot). The quantum state undergoes dissipation and cannot bounce back.}\label{4}
\end{figure}

In terms of the stationary phase approximation the solutions so far
obtained are for the center of a (gaussian) quantum state moving
along classical paths. Thus, in most cases the complete picture of
the collapse followed by an expansion is set to occur given
decoherence is absent. Our two final examples deal with this
possibility. The first case of Fig.\ref{4} shows a collapsing
solution obtained for boundary conditions with $\zeta_i>0$ near the
cutoff. Because in the commutative regime (dashed) nothing prevents
the system from collapsing all the way down to Planckian scales the
system will eventually enter the noncommutative regime with boundary
values at the cutoff (dot) compatible with a noncommutative
evolution (solid) that, just as the previous solutions, avoids
singularities and also displays the irregular oscillatory behavior
which is the strong indicator of noncommutative effects taking
place. As the center of the quantum state remains oscillating within
Planck length scales it can be said the state has dissipated due to
decoherence.

\begin{figure}[H]
  \centering
  \includegraphics[width=12cm]{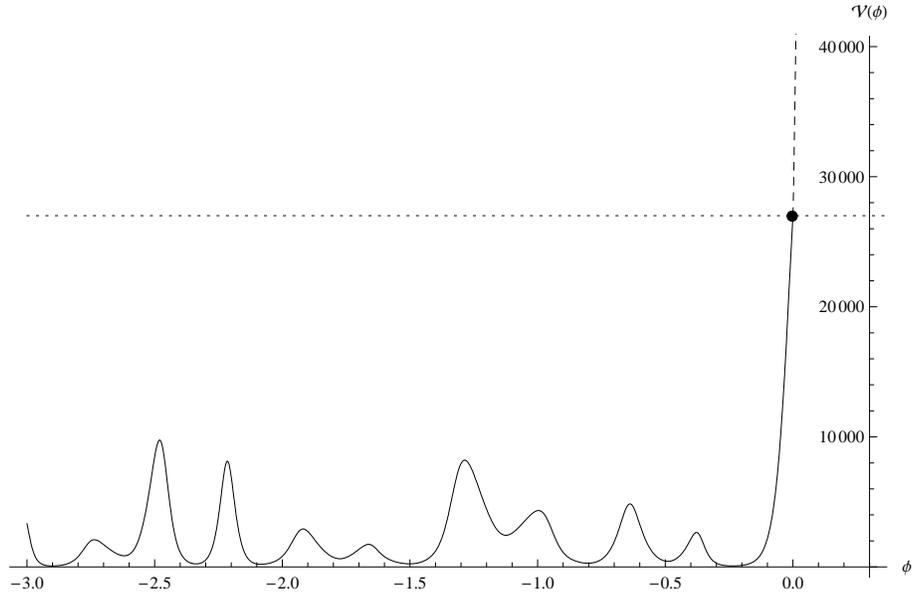}\\
  \caption{Expanding solution for $\varepsilon_i = 0.031 (\lambda_P)^{-1}$ . For a fixed cutoff value $L_i= 30\lambda_P $
 the noncommutative regime solution (solid) expands from decoherence reaching
 the cutoff region (dotted) following a commutative evolution algebra (dashed) compatible with the boundary values (dot).}\label{3}
\end{figure}

Time reversing the previous scenario would lead to a situation where the quantum state evolves from
decoherence to an expansion. Fig.\ref{3} corresponds to the numerical solution for this case characterized by $\zeta_i<0$ near
the cutoff. Once again the noncommutativity driven oscillations of
irregular amplitudes are noted before the system reaches the commutative regime
by means of the noncommutative force term discussed previously. Above
the cutoff the volume evolves according to (\ref{scf9}-\ref{scf11}) with
boundary values at the cutoff (dot).

\section{Conclusions}

In this article we approach Quantum Cosmology from the
point of view of a minisuperspace of a theory of Quantum Gravity .
We employ in particular the noncommutative ${\mathcal
C}^\star$-algebra $\frak A$ outlined in Sections II and III which
provides a well founded mathematical structure for introducing the
concept of noncommutativity, from the point of view of an
operational impossibility of measurement at distances smaller than a
few orders of the Planck length. This approach also allowed us to
relate the ${\mathcal C}^\star$-algebra formulation to some aspects
of the Loop Quantum Cosmology, as mentioned in Section III as well
as in the discussion of the asymptotics and numerics in Sections VI
and VII. In fact, taking $\varepsilon_i \to 0$ in (\ref{Lambda2})
reduces our noncommutative ${\mathcal C}^\star$-subalgebra of $\frak
A$ to a subalgebra of commutative $\hat U_i$'s which, together with
(\ref{vmult}), would lead to essentially the same results as those
contained in Ref.\cite{ash1}. Moreover, when considering the
$\varepsilon_i$ as scale factors and acted by a test function of
compact support which regularizes them, we have that the limit
$\varepsilon_i \to 0$ decouples $\varepsilon_i$ from $\mu_i$ in
(\ref{rel3}) and (\ref{rel4}). Hence, as shown in (\ref{scf11c}),
the $\mu_i$ are always of the order of a Planck length. This implies
that the granularity attributed to space in LQG is induced in our
formalism due to noncommutativity. Also the LQG variables involve
the concept of holonomies. But holonomies are naturally understood
in the theory of principal fiber bundles as integrals of connections
between two fibers. Although the trajectories resulting from these
integrals are not necessarily closed in the bundle space, they are
when projected to the base space. This would suggest the idea of the
loops. However, there is nothing in classical differential geometry
that says that the loops cannot have infinitesimal radii when the
fibers over base space are infinitesimally close. To have a minimal
radius one has to assume a discrete underpinning the continuum of
base space, which accounts for the "granularity" of space in LQG and
is reflected in the introduction of non-piecewise parameters of the
Heisenberg-Weyl group in order to avoid the implications of the
Stone-von Neumann Theorem. Thus "granularity" in LQG corresponds to
noncommutativity in our formulation. Moreover, connections (gauge
fields) are, according to Connes' Noncommuative Geometry, a
consequence of noncommutativity \cite{connes3}, so all this
therefore suggests its underlying presence
in the three main approaches mentioned in the Introduction.  \\

In Sections IV-VII the quantum collapse of a Bianchi I Universe was
studied in the context of noncommutative geometry. The
noncommutativity of the space variables (the axes of the Bianchi
Universe) was taken into account in a consistent way by representing
them in terms of the twisted discrete translation group algebra of
Sec.II. This representation is then used to construct the transition
amplitude by using the Feynman integral formalism, which was shown
to be dominated by an effective action that provides a new set of
equations that resulted to have a new dynamical behavior that took
into account the effect of the noncommutativity. It was shown
asymptotically and numerically in a generic case that the
noncommutativity induces an oscillatory motion of the volume due to
the nontrivial evolution of the action variables which are constant
for reticular space commutative theories. We thus have that the
dynamical effects of noncommutative produce an oscillatory behavior
of the volume in the region of the quantum bounce of reticular space
commutative theories. It will be interesting to study if these
oscillations in a full quantum field theory with spatial degrees of
freedom can be indeed interpreted as a topological change. The
differences mentioned above between our formalism and LQC lead to
some additional physical implications which result from our GNS
construction of the kinematic Hilbert space. The basic point being
that the reticulation induced on the arguments of the Hilbert space
contain at each point a tower of states, generated by the
consistency conditions required between the twisted translations
produced the unitaries $\hat U$'s and the translations due to the
$\hat V$'s. This implies that our reticulation induced by
noncommutativity is not the same as that in Ref.\cite{ash2} and
allows us to have, within the cosmology, a mechanism which could
prevent that all the fluctuations in our Bianchi I universe could
grow, thus avoiding to have a bounce at low matter densities. This
fundamental characteristic is obtained only in the improved version
of the polymeric cosmology of LQC, while in our case it occurs
naturally because of the way noncommutativity was implemented.
Moreover, in spite of the persistent difficulties inherent to this
field of research to obtain experimental information, we could hope
that phenomena lying in the interface of general relativity and
quantum physics, such as those involving quantum entanglement and
quantum coherence and which may be accessible to the experiment in
the near-term future, could provide further theoretical insights to
a full quantum theory of gravitation. This is suggested by the study
of noncommutativity in a simpler problem \cite{tim} where it was
shown that depending on the width of the wave packet of a coherent
state one could go from the commutative regime for wide packets to
the noncommutative regime for narrow packets. To perform this
evolution one needs to find a consistent analogue of the
Schr\"odinger equation in the noncommutative regime, and solve this
equation asymptotically as well as numerically in order to
understand this transition. This is currently under study.

\section{Acknowledgment}
The authors are grateful to Prof. Michael Ryan for many stimulating
discussions on the subject and to Prof. Noel Smyth, from the
Department of Mathematics and Statistics, The King's Buildings,
University of Edinburgh, for valuable help with some preliminary
numerical analysis of the system. This work was supported in part by
CONACyT Project  UA7899-F (M.R., A.A.M.) and DGAPA grant
UNAM-IN109013 (J.D.V.).

\end{document}